\documentclass[fleqn,usenatbib]{mnras}

\usepackage{newtxtext,newtxmath}

\usepackage[T1]{fontenc}

\usepackage{subfigure}

\DeclareRobustCommand{\VAN}[3]{#2}
\let\VANthebibliography\thebibliography
\def\thebibliography{\DeclareRobustCommand{\VAN}[3]{##3}\VANthebibliography}

\usepackage{graphicx}	
\usepackage{amsmath}	

\usepackage[normalem]{ulem}

\newcommand\chng{}

\title[Supermassive black holes in a mass-limited galaxy sample]{Supermassive black holes in a mass-limited galaxy sample}

\author[Z. Byrne et al.]{
Zachary Byrne$^{1}$, 
Michael J.\ Drinkwater$^{1}$,
Holger Baumgardt$^{1}$,
David Blyth$^{1}$,
Patrick {C{\^o}t{\'e}}$^{2}$,
\newauthor Nora L\"{u}etzgendorf$^{3}$,
Chelsea Spengler$^{4}$,
Laura Ferrarese$^{2}$,
Smriti Mahajan$^{5}$,
Joel Pfeffer$^{6}$,
Sarah Sweet$^{1,7}$\\
$^{1}$School of Mathematics and Physics, The University of Queensland, Brisbane, Queensland, 4072, Australia\\
$^{2}$National Research Council of Canada, Herzberg Astronomy and Astrophysics Program, Victoria, BC V9E 2E7, Canada\\
$^{3}$ESA Space Telescope Science Institute, Baltimore, USA\\
$^{4}$Institute of Astrophysics, Pontificia Universidad Catolica de Chile, Santiago, Chile\\
$^{5}$Department of Physical Sciences, Indian Institute of Science Education and Research Mohali, Punjab, India\\
$^{6}$International Centre for Radio Astronomy Research, The University of Western Australia, Perth, Australia\\
$^{7}$ARC Centre of Excellence for All Sky Astrophysics in 3 Dimensions (ASTRO 3D)\\
}

\date{Accepted XXX. Received YYY; in original form ZZZ}

\pubyear{2021}
\begin{document}
\label{firstpage}
\pagerange{\pageref{firstpage}--\pageref{lastpage}}
\maketitle


\begin{abstract}
The observed scaling relations between supermassive black hole masses and their host galaxy properties indicate that supermassive black holes influence the evolution of galaxies. However, the scaling relations may be affected by selection biases. We propose to measure black hole masses in a mass-limited galaxy sample including all non-detections to inprove constraints on galaxy mass - black hole mass scaling relations and test for selection bias. We use high spatial resolution spectroscopy from the Keck and Gemini telescopes, and the Jeans Anisotropic Modelling method to measure black hole masses in early type galaxies from the Virgo Cluster. We present four new black hole masses and one upper limit in our mass-selected sample of galaxies of galaxy mass (1.0--3.2) $\times 10^{10} M_\odot$. This brings the total measured to 11 galaxies out of a full sample of 18 galaxies, allowing us to constrain scaling relations. We calculate a lower limit for the average black hole mass in our sample of {\chng $3.7 \times 10^{7} M_\odot$.} This is at an average galaxy stellar mass of $(1.81 \pm 0.14)\times 10^{10} M_\odot $ and an average bulge mass of $(1.31 \pm 0.15) \times 10^{10} M_\odot$. This lower limit shows that black hole masses in early type galaxies are not strongly affected by selection biases.


\end{abstract}




\section{Introduction}
\label{sec:intro}

Supermassive black holes (SMBHs) are thought to be ubiquitous in the centres of large galaxies. After the early detections of SMBHs in nearby galaxies \citep[][]{Kormendy1995,Richstone1998}, larger samples revealed surprisingly tight correlations (or `scaling relations') between the masses of the central SMBHs and the properties of their host galaxies \citep[reviewed by][]{Kormendy_2013}. The central SMBH masses correlate with the host galaxy luminosity \citep[e.g.][]{Dressler1989}, bulge mass \citep[e.g.][]{Magorrian1998} and velocity dispersion \citep[e.g.][]{Gebhardt00,Ferrarese2000}. 

The tightness of these scaling relations motivated investigations into the underpinning physical processes. Most of the suggestions are known as black hole feedback as they involve energy transfer from the central SBMH to the surrounding galaxies \citep[e.g.][]{SilkRees,King2003}. Some form of black hole feedback is now recognised as essential in simulations of galaxy formation to prevent too many stars forming \citep[e.g.][]{Schaye2015}. A more complex view of the scaling relations is now emerging, with additional processes \citep[][]{Kormendy_2013} such as multiple mergers making the correlations tighter in giant elliptical galaxies than smaller galaxies.

The black hole scaling relations were initially proposed as single power law functions \citep[e.g.][]{Magorrian1998,Gebhardt00}. This has been revised as larger samples have become available: \citet[][]{Graham_2014} argue that a double power law relation between SMBH mass and galaxy spheroid mass is required when lower-mass galaxies are included. Their relation has a turnover point at roughly $10^{10} M_\odot$ with a steeper power law slope (2.2) for the low mass galaxies than for the high-mass galaxies (1.0). There are several reasons a different relation might be expected for the low-mass galaxies -- as discussed by \citet[][]{Graham_2014}. First, the scaling relation may change according to galaxy type \citep[][]{Davis_2019,Kraj2017,Shankar_2019}. Second, it may be difficult to define bulge masses for the lower-mass galaxies that are consistent with bulge masses used in the scaling relations for high-mass galaxies \citep[][]{Saglia2016}. This is less of an issue if additional galaxy properties (velocity dispersion, density or radius)  are used to form {\it bivariate} scaling relations \citep[][]{Saglia2016,Bosch_2016}. 

A further issue, reported by \citet[][]{shankar16}, is that dynamical measurements of SMBHs in early type galaxies may be skewed towards higher SMBH masses. This arises from the observational requirement that the sphere of influence of the respective SMBHs must be resolved. \citet[][]{shankar16} model this bias and obtain the observed scaling relations from much steeper true relations. The bias increases to lower galaxy masses: at a galaxy stellar mass of $10^{10} M_\odot$ the observed relation has SMBHs an order of magnitude more massive than the true relation. 

In this paper we present a project to measure SMBH masses in a complete, mass-selected, sample of early-type galaxies. Our aim is to test the observational bias proposed by \citet[][]{shankar16} by measuring all the galaxies in our sample and including any non-detections in our estimate of the average SMBH mass. We choose a sample with galaxy masses near $10^{10} M_\odot$ as this is where \citet[][]{shankar16} predict the largest bias. This is also a galaxy mass at which it will be useful to test galaxy formation simulations \citep[e.g.\ fig 10 of][]{Schaye2015}.

We describe our sample selection and methodology in Section 2 and present our new SMBH mass measurements in Section 3. We discuss our results in the context of published scaling relations in Section 4. {\chng We also discuss possible biases from the axisymmetric assumption made in our dynamical method (Jeans) and analysis}. We have not yet measured all the galaxies in our sample, but we are able to estimate a lower limit to the average SMBH mass in our sample that is too large to be consistent with some published scaling relations.

\section{Method}
\label{sec:Method}
In this section we describe our approach to construct a representative sample of early-type galaxies and our methods to estimate black hole masses in these galaxies. We used imaging data from the Hubble Space Telescope (HST) and high spatial resolution spectroscopic data from the Gemini North and Keck telescopes.



\subsection{Sample Selection}
\label{sec:samplesel}

We constructed a mass-selected sample of early-type galaxies from galaxies observed by the Advanced Camera for Surveys Virgo Cluster Survey (ACSVCS) \citep[][]{acsvcsnuc} with mass estimates from the ATLAS 3D sample \citep[][]{atlas3dfirst}. 

We chose galaxies from the ACSVCS survey as this ensured the availability of the high spatial resolution (HST) images required to generate mass models of each galaxy (see Sec.~\ref{sec:mgefits}). The ACSVCS randomly selected early-type galaxies from the Virgo Cluster, so this requirement does not bias the sample in any way likely to affect black hole mass other than the early-type and possibly the cluster environment. 

The ATLAS 3D masses were calculated as follows by \citet[][]{atlas3d}. They used Sloan Digital Sky Survey \citep[][]{sdss} and Isaac Newton Telescope photometry to estimate the surface brightness profile of the galaxies, and integral field observations from the Spectrographic Areal Unit for Research on Optical Nebulae (SAURON) survey \citep[][]{sauron} for the galaxy kinematics. They then used dynamical models to calculate total mass-to-light ratios within one effective radius of each galaxy. The total ATLAS 3D mass (as shown in Table~\ref{tab:sample}) is the product of the mass-to-light ratio and the total luminosity. The median dark matter fraction in these central regions is only 13 per cent, so the ATLAS 3D total galaxy mass is close to the total stellar mass. We use a similar method to estimate SMBH masses, as described in Section \ref{sec:dynmodel}.

We selected galaxies in the ATLAS 3D mass range ($1- 3.2) \times 10^{10} M_\odot $. We chose this mass range to directly test for bias in the black hole scaling relation \citep[][]{shankar16} as well as covering the mass range examined by other studies. Our sample numbers 18 galaxies from the Virgo Cluster, as presented in Table~\ref{tab:sample}. Out of the sample of 18 mass-selected galaxies, we have observed 5 galaxies ourselves. The black hole measurements of 6 galaxies are publicly available (a combination of \citet[][]{krajonvicbh} and \citet[][]{Nowak2007}), and the remaining 7 galaxies are yet to be observed using appropriate instruments. We present our initial results from the observed 5 galaxies in this paper.


In summary, we have constructed a representative sample of 18 early-type galaxies, by selecting all galaxies observed by the ACSVCS survey within the total mass range of $(1- 3.2) \times 10^{10} M_\odot$, as given by ATLAS 3D \citep[][]{atlas3d}.

\begin{table*}
  \begin{center}
   \caption{Mass-selected galaxy sample with results}
   \label{tab:sample}
    \begin{tabular}{l|l|r||r}  
      \textbf{Galaxy} & \textbf{Source of spectroscopy} & \textbf{Galaxy Mass} & \textbf{Black hole mass} \\
       & Instrument and reference & ($M_\odot$) &($M_\odot$) \\
      \hline
      NGC 4262 & \textbf{Gemini North NIFS} & $3.0 \pm{0.2} \times 10^{10}$ & {\chng \boldmath$3.15 \pm{0.41} \times 10^8$}\\
      NGC 4339 & Gemini North NIFS \citep[][]{krajonvicbh} & $2.7 \pm{0.2} \times 10^{10} $ & $6.5 \pm{1.0} \times 10^7$ \\
      NGC 4377 & Not observed & $1.7 \pm{0.1} \times 10^{10}$   &\\
      NGC 4379 & \textbf{Gemini North NIFS} & $1.9 \pm{0.1} \times 10^{10}$ & {\chng \boldmath$7.07 \pm{0.36} \times 10^7$} \\
      NGC 4387 & Not observed & $1.5 \pm{0.1} \times 10^{10}$  & \\
      NGC 4434 & Gemini North NIFS \citep[][]{krajonvicbh} & $1.6 \pm{0.1} \times 10^{10} $ & $9.0 \pm{1.0} \times 10^7$ \\
      NGC 4452 & Not observed & $1.8 \pm{0.1} \times 10^{10}$   &\\
      NGC 4458 & \textbf{Gemini North NIFS} & $1.1 \pm{0.1} \times 10^{10}$ & {\chng \boldmath$6.51 \pm{0.45}\times 10^7$} \\
      NGC 4474 & Gemini North NIFS \citep[][]{krajonvicbh} & $1.5 \pm{0.1} \times 10^{10} $ & $<0.15 \times 10^7$ \\
      NGC 4476 & Not observed & $1.0 \pm{0.1} \times 10^{10}$ &  \\
      NGC 4483 & Not observed & $1.4 \pm{0.1} \times 10^{10}$ &  \\
      NGC 4486A & Very Large Telescope SINFONI \citep[][]{Nowak2007}&$1.6 \pm{0.1} \times 10^{10} $ & $1.25 \pm{0.75} \times 10^7$  \\
      NGC 4528 & Not observed & $1.3 \pm{0.1} \times 10^{10}$ &   \\
      NGC 4550 & \textbf{Keck OSIRIS} & $2.5 \pm{0.2} \times 10^{10}$ & {\chng \boldmath$1.14 \pm{0.37} \times 10^7$} \\
      NGC 4551 & Gemini North NIFS \citep[][]{krajonvicbh} & $1.8 \pm{0.1} \times 10^{10} $ & $<0.8  \times 10^7$ \\
      NGC 4578 & Gemini North NIFS \citep[][]{krajonvicbh} & $2.8  \pm{0.2}\times 10^{10}$  & $3.5 \pm{0.3} \times 10^7$  \\
      NGC 4612 & Not observed & $1.8 \pm{0.1} \times 10^{10}$   & \\
      NGC 4623 & \textbf{Gemini North NIFS} & $1.5 \pm{0.1} \times 10^{10} $ & {\chng \boldmath$<4.6\times 10^6$}  \\
    \end{tabular}
\\
    
  \end{center}

Note: Galaxy mass is the total dynamical ATLAS 3D masses, taken from \citet[][]{atlas3d}. Our results are in bold.
\end{table*}

\subsection{Imaging Data}
\label{sec:imaging}

Our dynamical models (Sec~\ref{sec:dynmodel}) rely on wide field images to create a mass profile of each galaxy. We used images from the ACSVCS \citep[][]{acsvcsfirst} survey which has observed all our sample galaxies with the Hubble Space Telescope. The 200 arcsec field of view of the ACS WFC camera is wide enough to fully sample each galaxy, whilst resolving the galaxy nuclei \citep[][]{acsvcsnuc}. The data reduction and image processing methodology for the ACSVCS are described by \citet[][]{Jordan2005}. We chose images with the F850LP filter as they are the most appropriate to trace stellar mass for galaxies. We converted the F850LP photometry to units of solar luminosity by first converting to the AB magnitude system using a zero-point of 24.871 mag and then using 4.50 mag for the absolute magnitude of the Sun in the F850LP filter \citep[][]{willmer2018}. We used distances for each galaxy (see Table~\ref{tab:ownbh}) from the compilation by \citet[][]{atlas3dfirst}.

\subsection{Spectroscopic Data}
\label{sec:spectro}
We measured the central kinematics of the galaxies with high spatial resolution infrared integral field unit (IFU) spectra from the Gemini and Keck telescopes. We used the adaptive optics systems with laser guide stars to deliver resolutions close to the diffraction limits of each telescope (half-width half-maxima of 0.022 and 0.027 arcsec for Keck and Gemini respectively). In each case we used the galaxy nucleus for the first order tip-tilt corrections. The laser system was unavailable for our 2017 Gemini observations, so these were made in conditions of very good natural seeing (FWHM $<$ 0.25 arcsec). Table \ref{tab:specobs} presents the details of our IFU spectroscopic observations.


We observed four of our galaxies, NGC4458, NGC4623, NGC4379 and NGC4262, using the Near-Infrared Integral Field Spectrometer  \citep[NIFS, ][]{nifspaper} on the Gemini-North telescope. NIFS includes the ALTtitude conjugate Adaptive optics for the InfraRed system \citep[ALTAIR, ][]{nifspaper}.  We observed in the K band (2.0--2.4 microns) at a spectral resolution of 5300, in 0.1 arcsec $\times$ 0.04 arcsec spaxels over a 3 arcsec field of view. We followed the recommended data reduction as outlined in the NIFS handbook \citep[][]{nifspaper} to produce final data cubes with a spatial sampling of 0.05 arcsec $\times$ 0.05 arcsec.

We observed our other galaxy in the sample, NGC4550, using the Keck telescope. We used Optical Spectrograph and InfraRed Imager System \citep[OSIRIS, ][]{Keck}. We used the Kn5 filter (2.29--2.41 microns) with a spectral resolution of 3900, and observed the inner region in  0.02 arcsec $\times$ 0.02 arcsec spaxels over a  0.6 arcsec $\times$ 1.2 arcsec field of view. We followed the data reduction guidelines as outlined in the OSIRIS handbook \citep[][]{Keck}. We made one modification to the standard approach, changing the combine method to a median average in order to maximise signal to noise. 

We estimated the spatial resolution of our IFU data by comparison with the corresponding HST images. For each galaxy, we fitted a multiple-Gaussian model to the central region of the HST images using the MGE package \citep[][]{MGE_cap}. This process allows for the HST PSF. We then summed each IFU spectral cube in the spectral direction to create a 2-dimensional galaxy image which was the convolution of the galaxy with the IFU PSF. We then created models of the IFU galaxy images by convolving images of the MGE fits with a double-Gaussian model of the IFU PSF. The IFU was modelled by three parameters: the widths of the two Gaussians and their relative normalisaion. We found the best fitting values by $\chi^2$ minimisation of the difference between the model and the observed IFU galaxy image along the major axis of the galaxy. The best values are given in Table~\ref{tab:specobs}. This approach and the results are similar to those of \citet{krajonvicbh}.

\begin{table*}
    \centering
    \caption{Spectroscopic observations}
    \label{tab:specobs}
    \begin{tabular}{l|r|l|l|r|r|r|r|l}
         \textbf{Galaxy} & \textbf{Date} & \textbf{PID} & \textbf{Instrument} & \textbf{Exp}& \textbf{Mode} &\textbf{$W_1$}&\textbf{$W_2$}&\textbf{$n_1$}  \\
             &     &     &     & (h) & & \multicolumn{2}{|c|}{(FWHM, arcsec)}&  \\
         \hline
         NGC 4262 & 2017-03-07 & GN-2017A-Q-22 & Gemini/NIFS & 2.0  & natural& 0.21  & 1.18 & 0.64  \\ 
         NGC 4379 & 2016-04-26 & GN-2016A-Q-35 & Gemini/NIFS & 2.3  & AO+LGS & 0.15  & 0.67 & 0.69 \\  
         NGC 4458 & 2016-04-23 & GN-2016A-Q-35 & Gemini/NIFS & 1.75 & AO+LGS & 0.23  & 0.57 & 0.95  \\ 
         NGC 4550 & 2016-05-15 & Z039OL        & Keck/OSIRIS & 2.0  & AO+LGS & 0.065 & 0.88 & 0.83 \\ 
         NGC 4623 & 2017-04-10 & GN-2017A-Q-22 & Gemini/NIFS & 1.0  & natural& 0.22  & 0.98 & 0.73 \\  
    \end{tabular}
\\
Notes: exposure times include sky measurements which were typically 40\% of the total time. $W_1$ and $W_2$ are the widths (full width half-maximum, in arc seconds) of double-Gaussian fits to the point spread function of each observation. $n_1$ and $1-n_1$ are the respective intensities of the two Gaussians. 
\end{table*}





\subsection{Observed Kinematics}
Measuring the kinematics of the galaxy's inner regions is necessary to dynamically model the black hole's influence. We analysed the spectroscopic data to measure the central kinematics of each galaxy as follows.

First, we calculated an average spectrum for each galaxy by averaging the observed IFU spectra. We then compared this average galaxy spectrum to templates from the MILES stellar library \citep[][]{MILES}. We used the Penalised Pixel-Fitting program {\chng \citep[\textsc{pPXF}, ][]{Cappellari2004,ppxf}} to find the best fitting combination of templates from the stellar library to represent the average galaxy stellar population. We combined this combination of templates to form a master template spectrum (not smoothed by the velocity dispersion) for the galaxy. We used Voronoi binning \citep[][]{Voronicap} to average groups of adjacent spaxels, {\chng at the highest required signal-to-noise ratio that still resolved the central regions of each galaxy. The signal-to-noise limit was 100 or more for all galaxies except NGC 4550 which required a limit of 20.} We then compared the average spectrum in each Voronoi bin to the master template spectrum using \textsc{pPXF} \citep[][]{ppxf} to calculate the velocity and velocity dispersion of that region of the galaxy. {\chng In a few cases the binned spectra had very large noise spikes which we masked to avoid spurious kinematic measurements.}

\begin{figure*}
\begin{center}
\includegraphics[scale = 0.5] {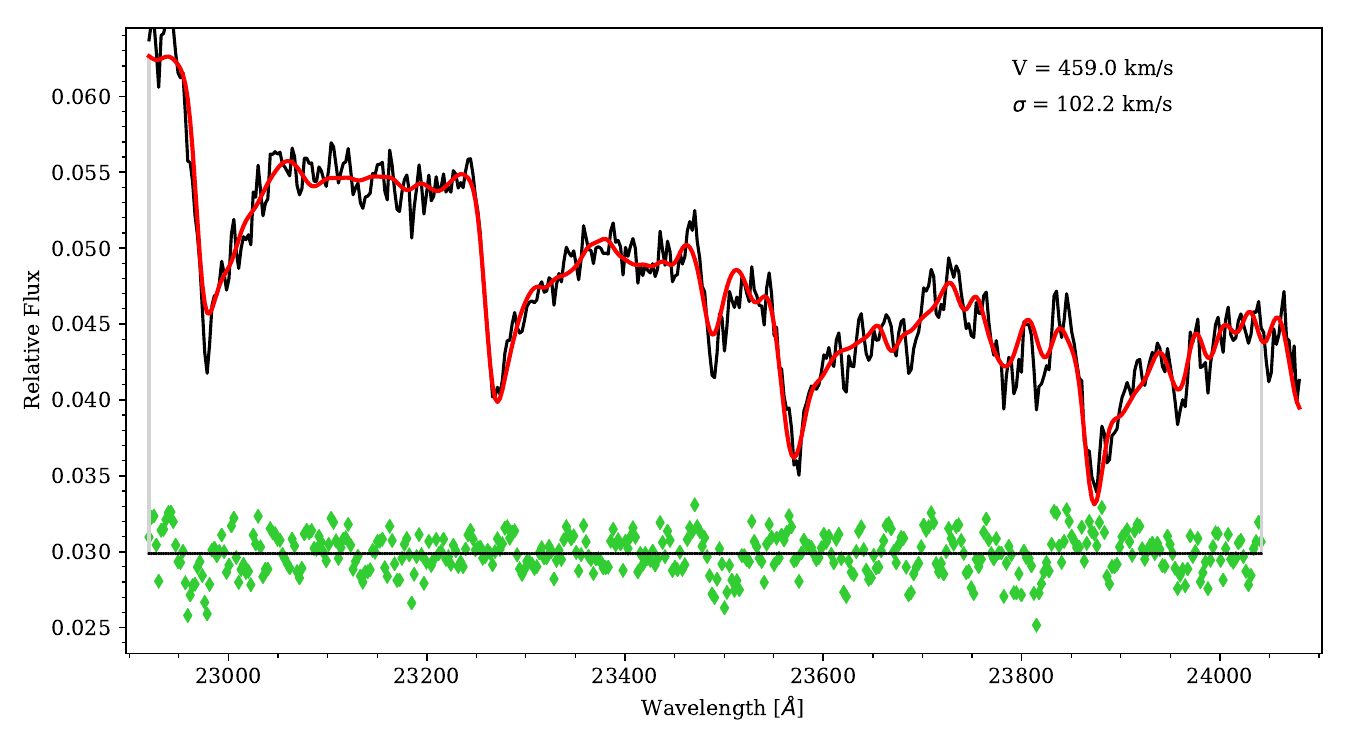}

 \caption{An example of the spectral fitting we use to measure velocity and velocity dispersion. The observed spectrum (black) is compared to our master template spectrum (red) after velocity dispersion smoothing and redshifting, and the deviation is shown in green below the spectra.\label{fig-specfit}}
 
 \end{center}
\end{figure*}

We show an example of how we measured the velocity and velocity dispersion in Fig.~\ref{fig-specfit} and we show the resulting velocity and velocity dispersion maps in the Voronoi bins in Fig.~\ref{fig:vormap}




\subsection{Dynamical Modelling}
\label{sec:dynmodel}
\label{sec:mgefits} 

Finally, we created dynamical models of each galaxy and compared them to the the observed central kinematics to estimate the mass of any SMBH. 

We used the optical imaging (Sec.~\ref{sec:imaging}) to constrain a large-scale mass model of each host galaxy. We used the Multi-Gaussian Expansion method \citep[MGE, ][]{mgediv} to model the stellar surface brightness distribution of the galaxy as a sum of two-dimensional concentric Gaussians. The MGE models are corrected for the PSF of the imaging data used. We implemented the MGE method by using the \textsc{MGEfit} Python package \citep[][]{MGE_cap}.  
 
Each MGE fit consisted of 7-10 Gaussian components. These components were aligned with the major photometric axis of the galaxy and centred the galaxy core, but their intensities, widths and ellipticities varied. We limited the region used (to 3\% of the image in the find\_galaxy routine of \textsc{MGEfit}) to estimate the central position (and major axis) of each galaxy to ensure that the MGE fits were centred on the cores of each galaxy. 

\begin{figure*}
\centering
  \includegraphics[width=.48\linewidth]{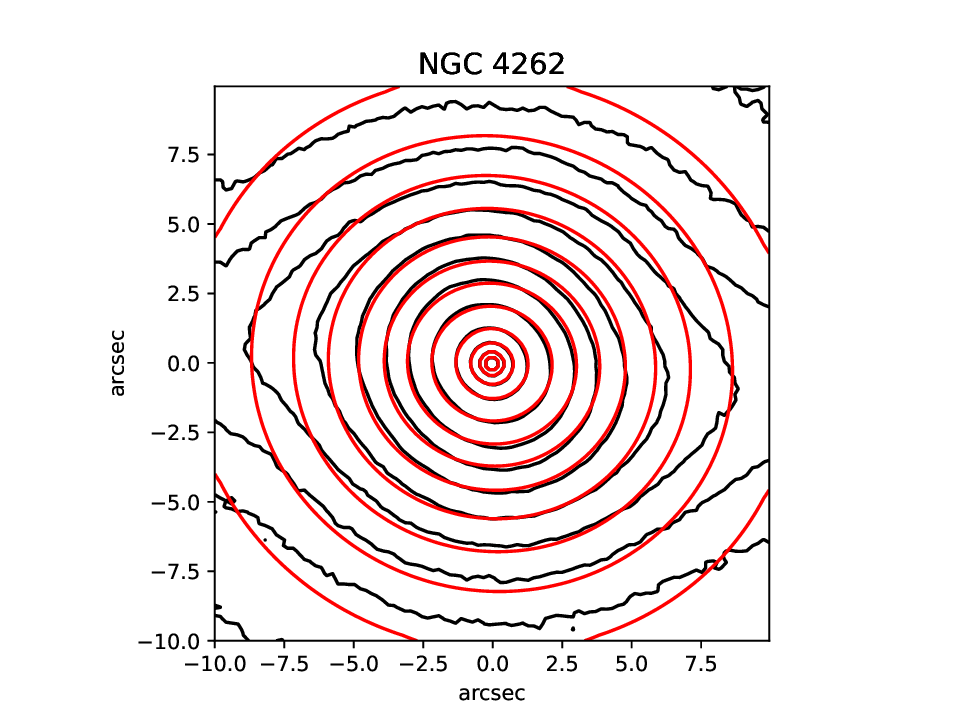}
  \includegraphics[width=.48\linewidth]{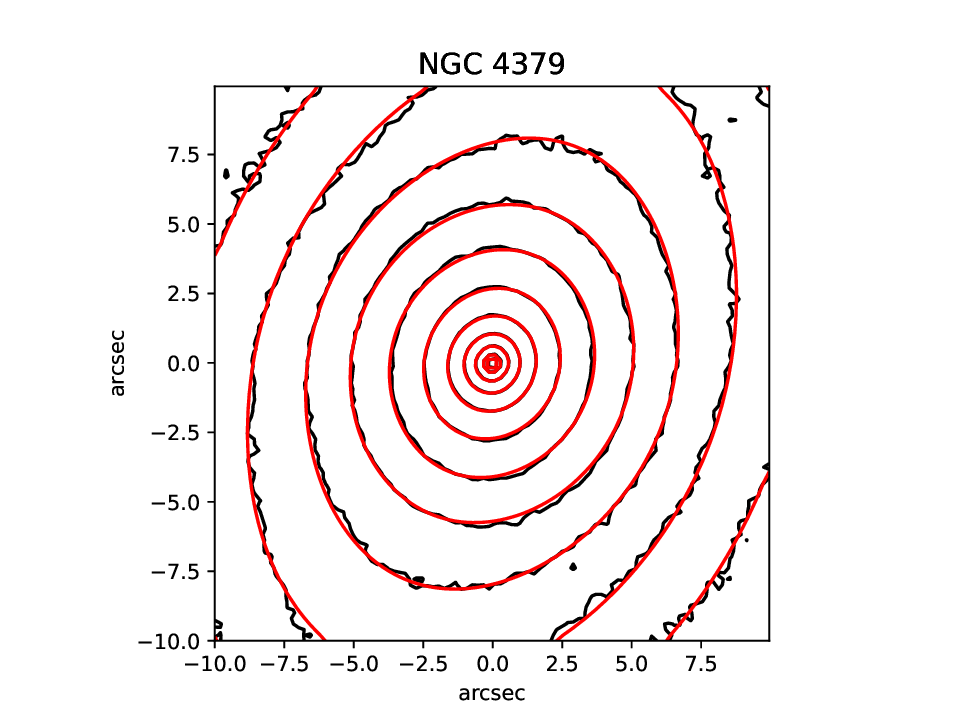}
  \includegraphics[width=.48\linewidth]{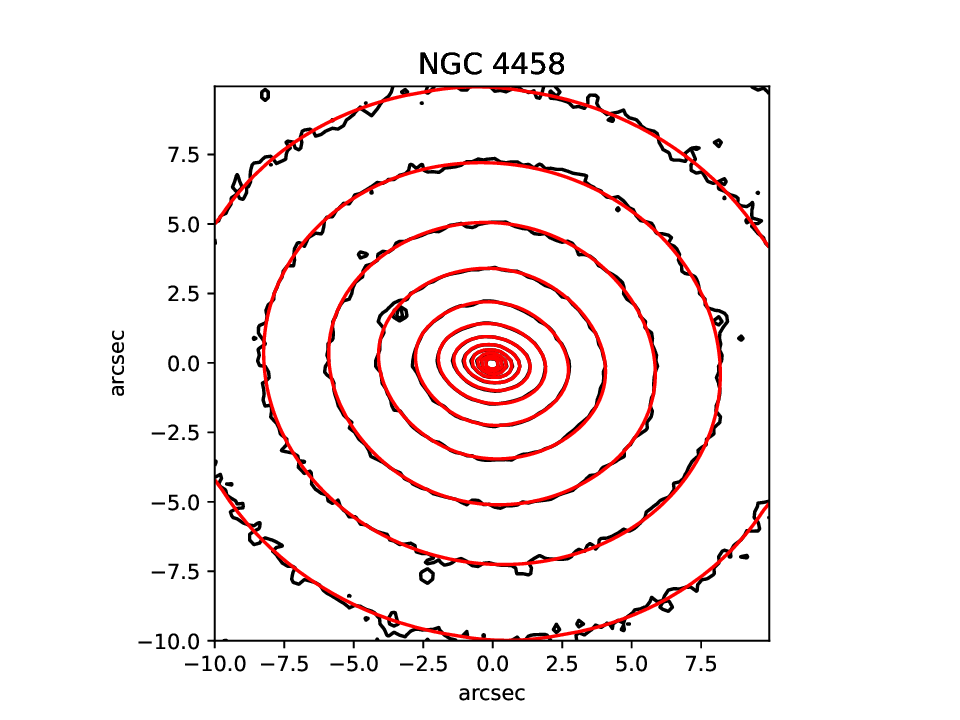}
  \includegraphics[width=.48\linewidth]{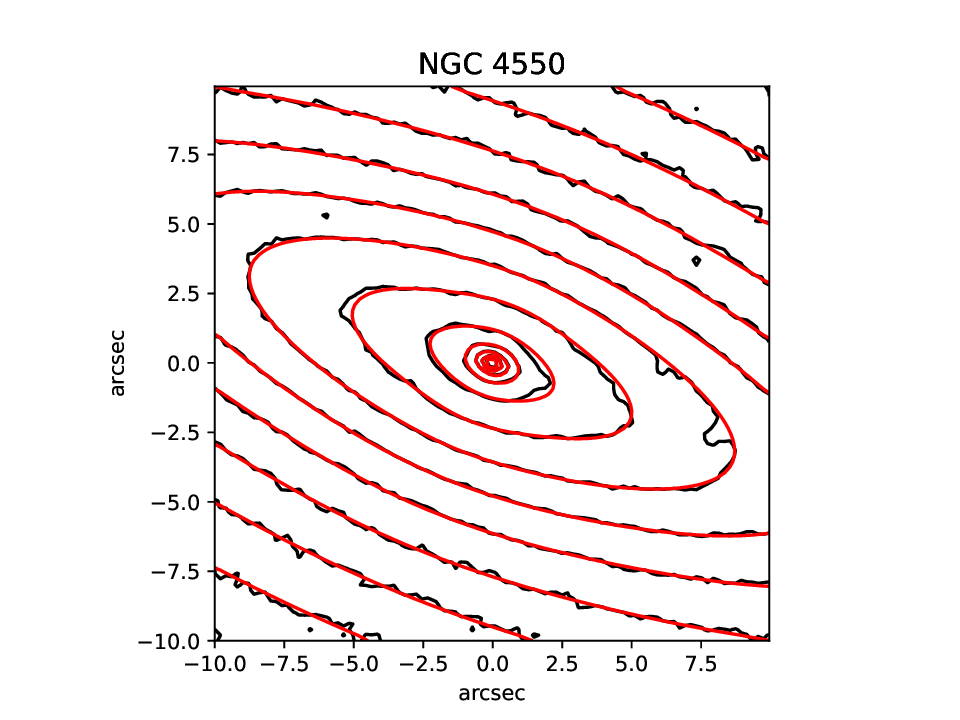}
  \includegraphics[width=.48\linewidth]{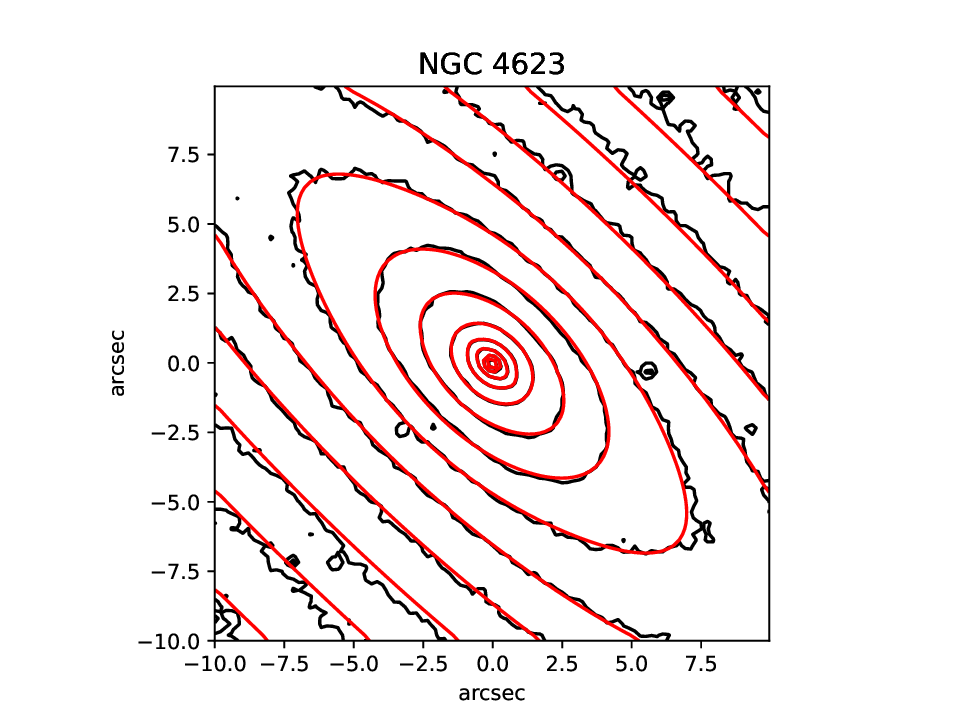}
   \caption{\label{fig:mge} Models of the surface brightness distributions of each galaxy compared to the observed HST images. The contours of the MGE models, convolved with the image PSF, are plotted in red and the corresponding contours of the original images are shown in black. }
  
\end{figure*}

We present the MGE fits for the galaxies in Fig. \ref{fig:mge} as contours of the fits and the original images in the central regions of each galaxy. As can be seen, the MGE fits reproduce the central image contours of each galaxy very well. In some cases, the outer regions (not shown) were not such good fits, but the outermost contours (10 magnitudes per arcsec$^2$ fainter than the peak values) of the models all matched the observations to within 12\% relative error in radius.



 


Starting with the structural (MGE) models of the host galaxies, we then modelled the kinematics for comparison to the  observed kinematics to determine if additional black hole contributions were required. We used the Jeans Anisotropic Modelling method \citep[as implemented in the \textsc{JamPy} Python package, ][]{JAMs} to solve the Jeans equations for the motion of stars in a gravitational field. {\chng  The \textsc{JamPy} package uses \textsc{MGEfit} structural models to construct dynamical models of galaxies and predict the resulting kinematic properties. The models depend on four free parameters: the inclination,} the mass of any central SMBH ($M_{bh}$), the anisotropy parameter ($\beta = 1 - \sigma_z^2/\sigma_R^2$ where $\sigma_z$ and $\sigma_R$ are the velocity dispersions along the cylindrical coordinates $z$ and $R$), and the mass-to-light ratio ($M/L$) of the structural model \citep[][]{MGE_cap}. The model predicts the observed velocity dispersion distribution in the central region of the galaxy allowing for the point spread function of the IFU spectroscopic observations.

{\chng 

We found the best parameters by $\chi^2$ minimisation of the difference between the model and observed velocity dispersion maps using an adaptive Metropolis algorithm \citep{Haario2001} \citep[as implemented the \textsc{adamet} Python package by][]{atlas3d}. We tested a wide range of each of the parameters for all galaxies: $-1<\beta<1$, $0<M_{bh}<10^9 M_\odot$, and $0<M/L<10$.  We varied inclination $i$ between a maximum of 90 degrees and a minimum possible value $i_{\rm min}$ determined by the smallest axial ratio $q_{\rm min}$ of the observed MGE components according to $\cos^2(i_{\rm min})= q^2_{\rm min}$.

One galaxy, NGC 4262, has a prominent bar. This results in one of its MGE components having a very low axial ratio ($q=0.55$), all other components being close to circular ($q>0.94$). To avoid an unphysical limit on the inclination of this face-on galaxy, we refit requiring $q>0.94$ for all MGE components. The MGE fit was still a good match  to the inner (see Fig.~\ref{fig:mge}) and outer contours and the total MGE luminosity changed by only 1 per cent.}

\section{Results}

In this Section we describe the analysis of our kinematic data to estimate masses of any detected black holes. 

We present two dimensional maps of the projected kinematics of our galaxies in Fig.~\ref{fig:vormap}. For each galaxy, the left hand panel shows the velocity map, and the right hand panel the velocity dispersion map. All of the galaxies show some evidence of systematic rotation in the velocity maps. In each galaxy, except NGC 4623, there is clearly an increase in velocity dispersion in the central regions of the galaxy.

\newcommand\vorwidth{0.65}
\newcommand\vorgap{-7mm}
\begin{figure*}
  \includegraphics[width=\vorwidth\linewidth]{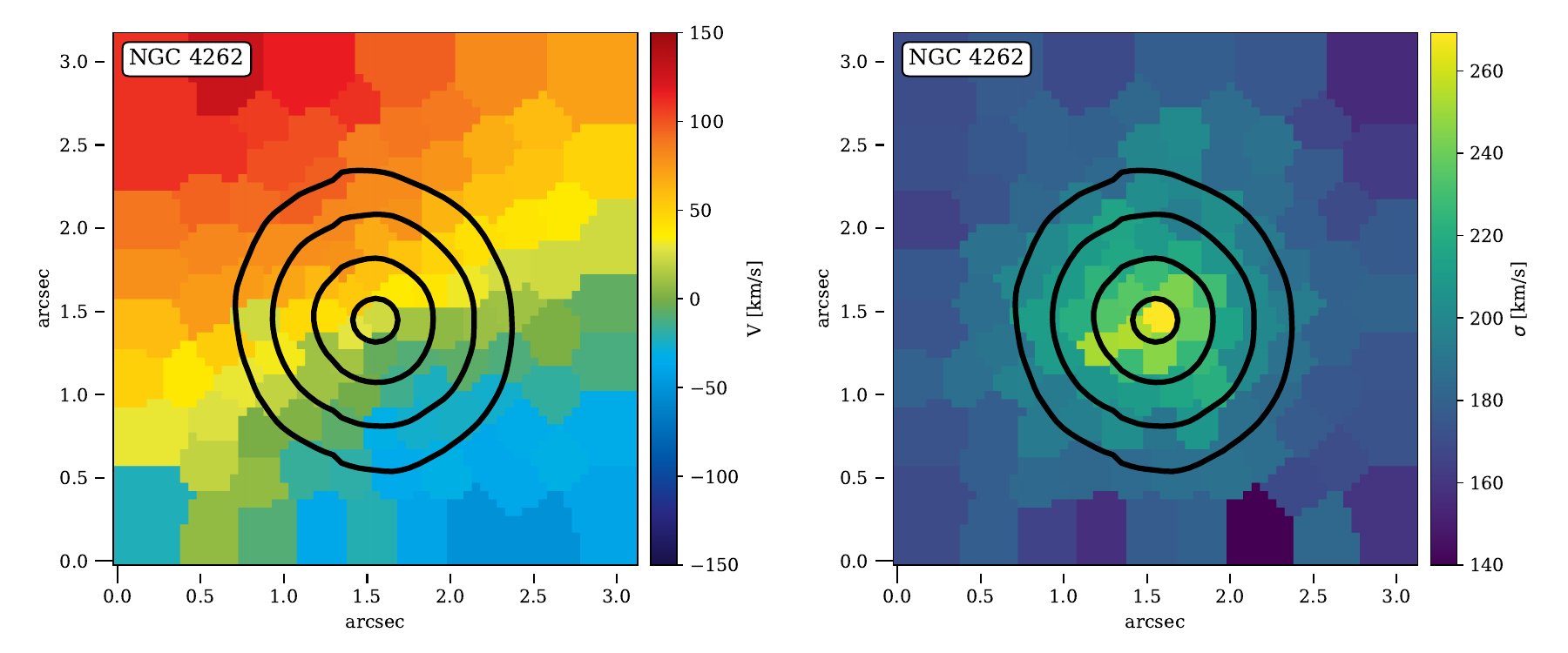}
  \includegraphics[width=\vorwidth\linewidth]{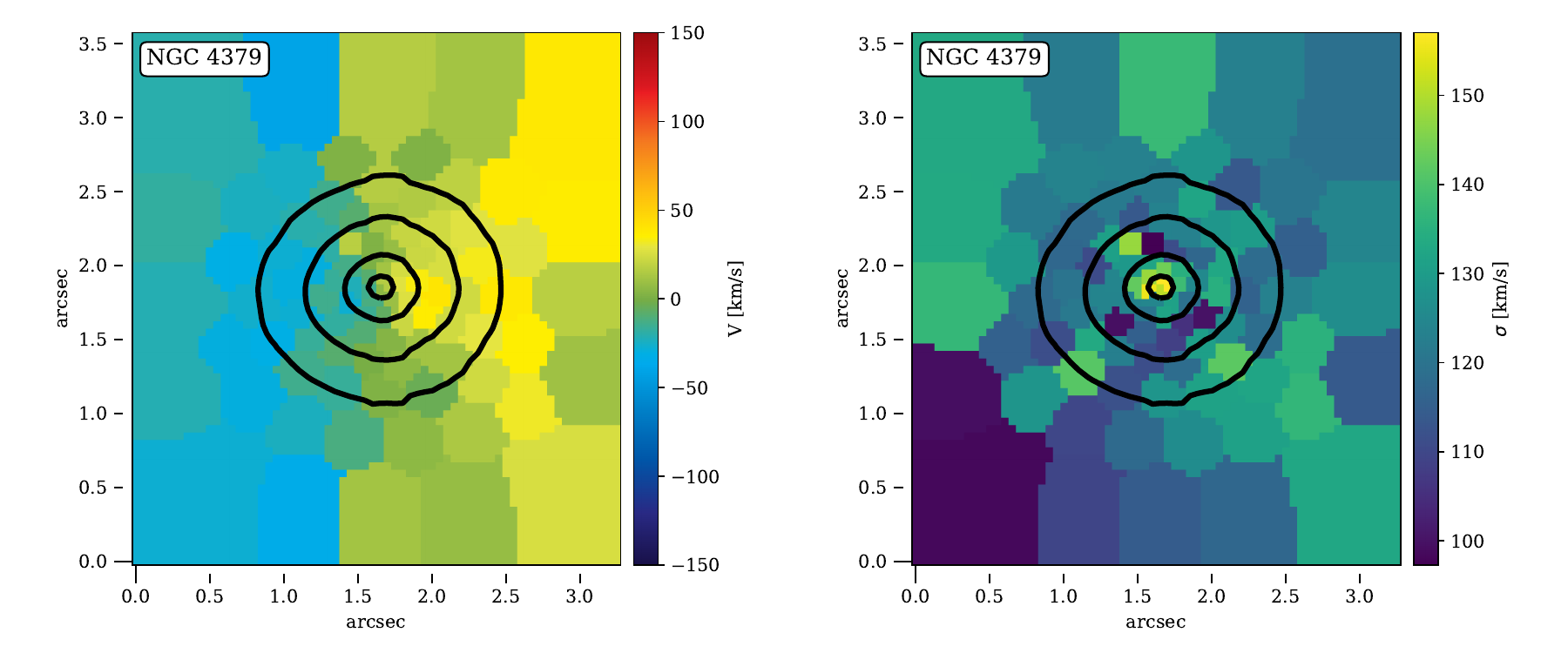}
  \vspace*{-8mm}
  \includegraphics[width=\vorwidth\linewidth]{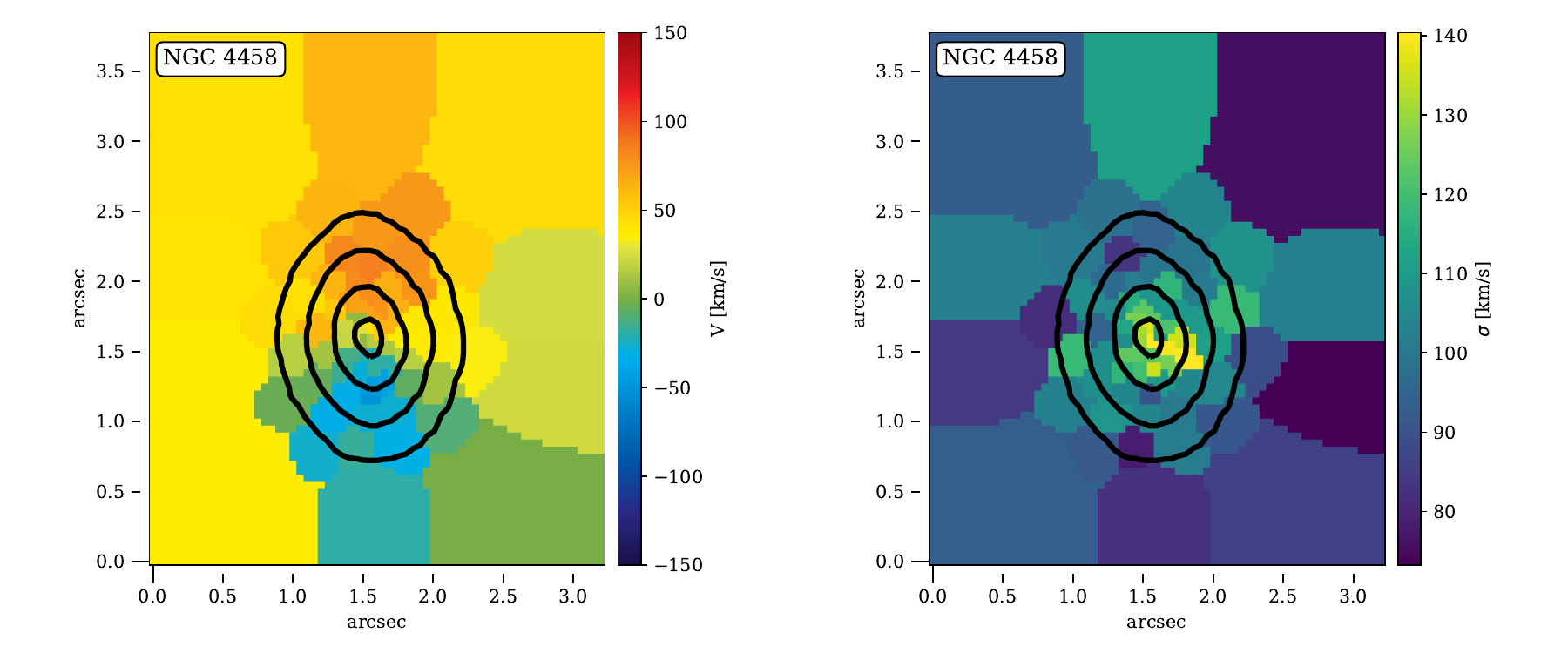}
  \vspace*{-8mm}
  \includegraphics[width=\vorwidth\linewidth]{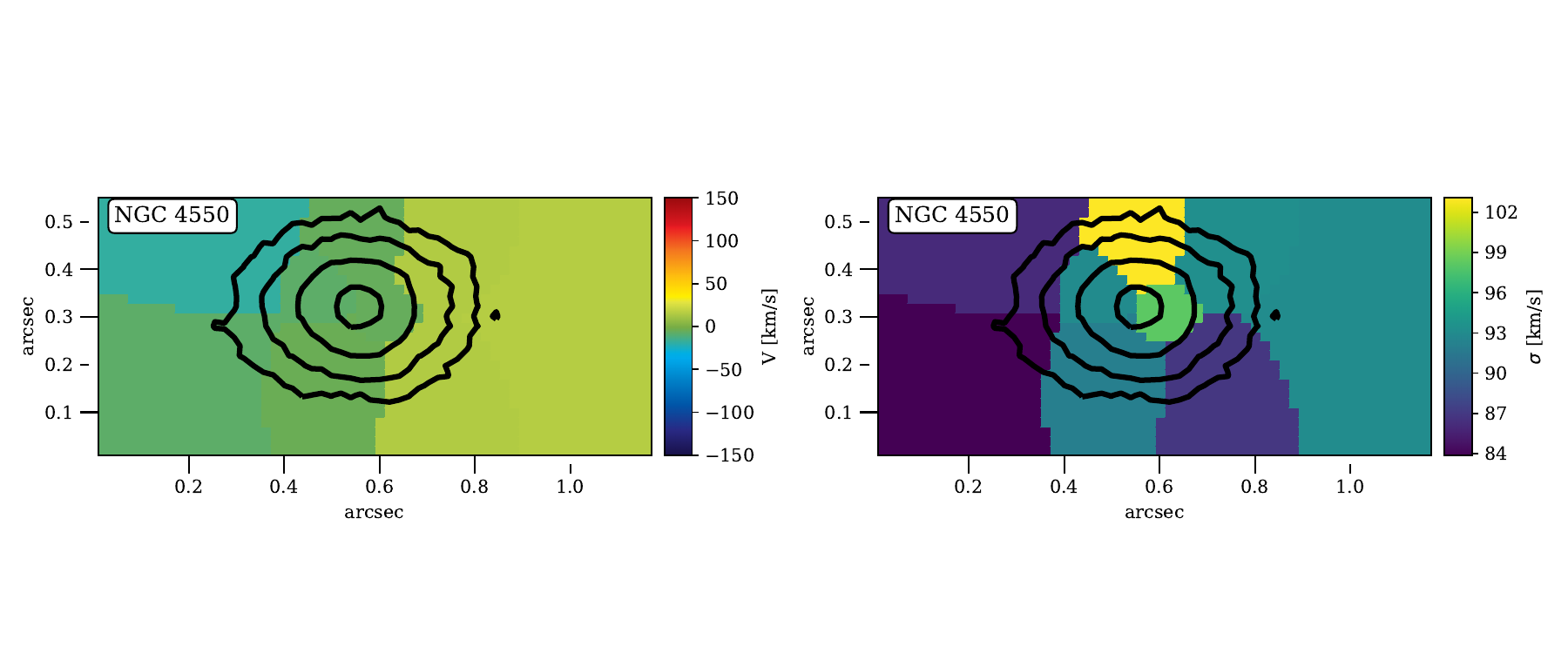}

  \includegraphics[width=\vorwidth\linewidth]{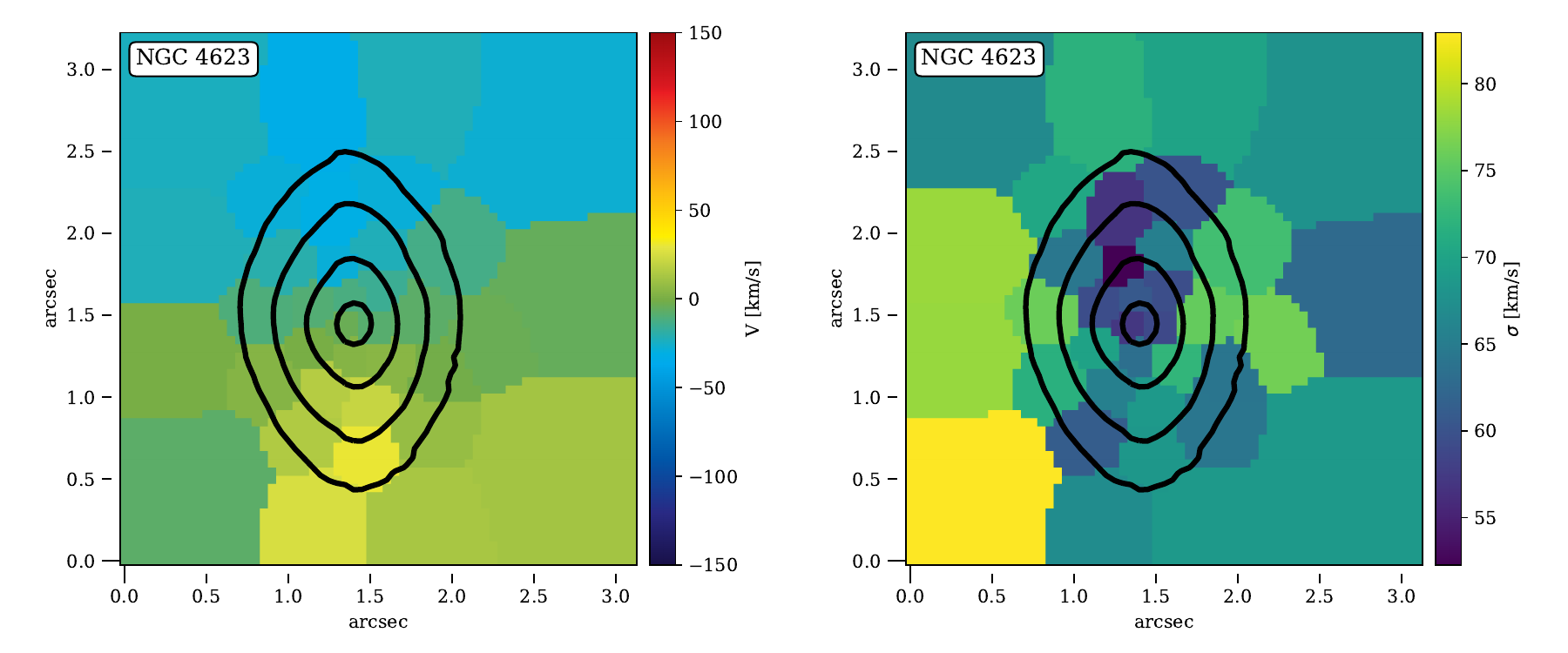}
  \caption{\label{fig:vormap}Velocity (left) and velocity dispersion (right) maps for each galaxy in Voronoi bins. The spaxel size is {\chng 0.05 arcsec} for all galaxies except for NGC 4550, which is 0.02 arcsec. The black contours show the total intensity at levels of 80, 40, 20 and 10 per cent of the maximum. The raised velocity dispersion at the centre of most galaxies is associated with a black hole detection. }
\end{figure*}

\begin{figure*}
\begin{center}
\includegraphics[scale = 0.5]{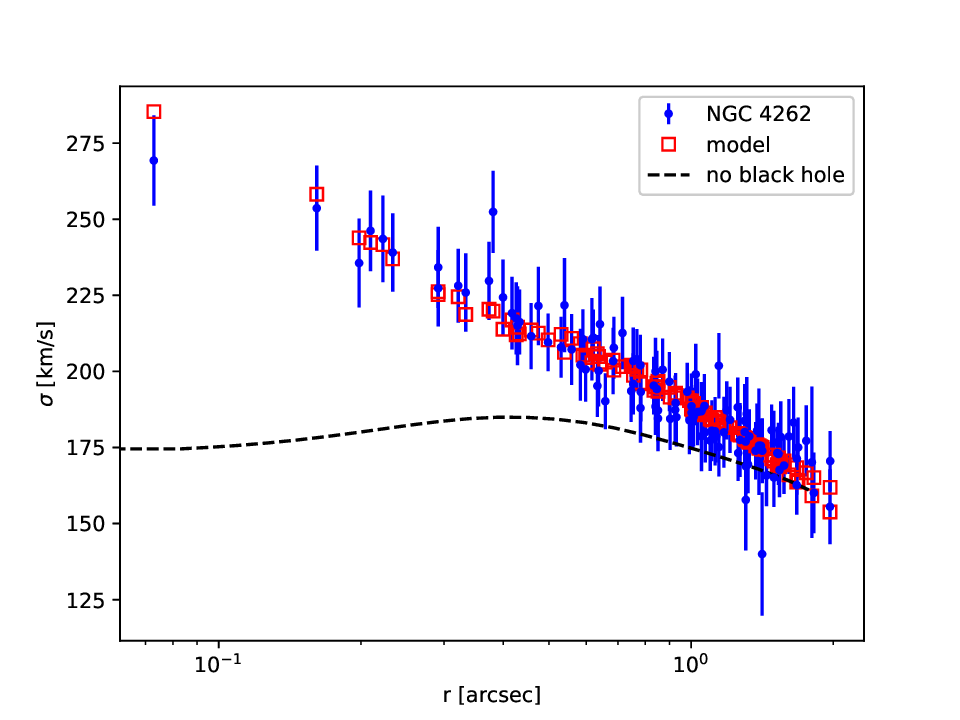}
\includegraphics[scale = 0.5]{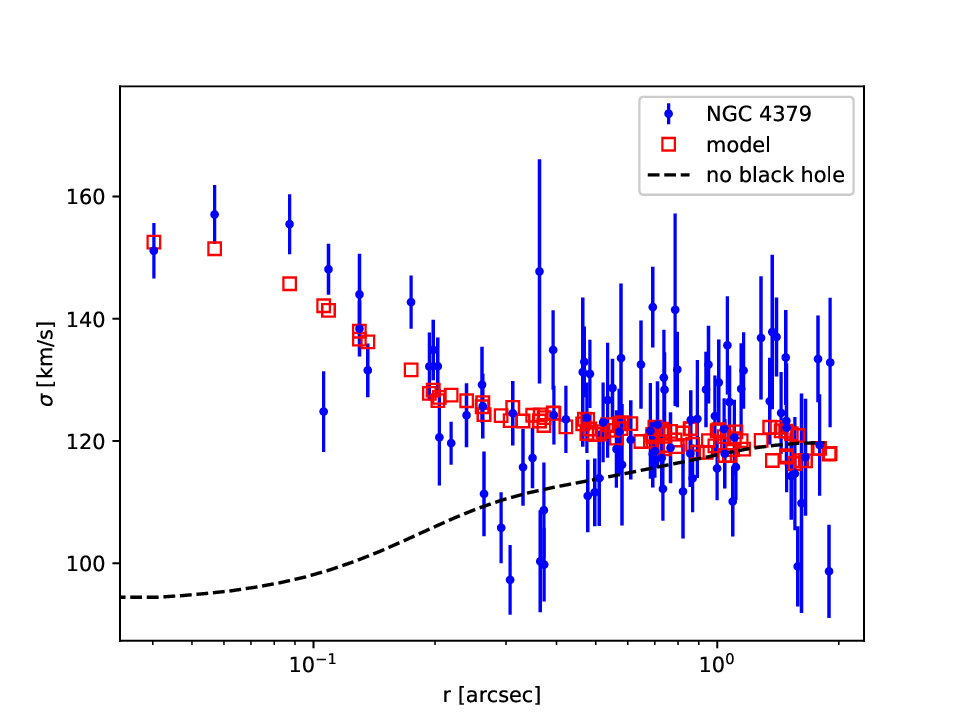}
\includegraphics[scale = 0.5]{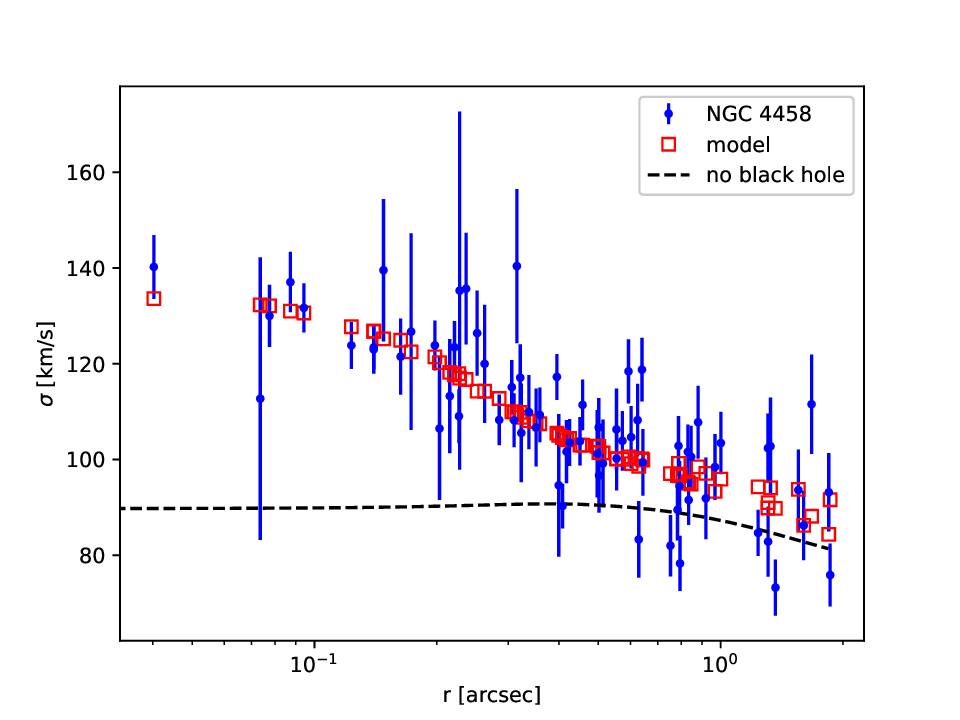}
\includegraphics[scale = 0.5]{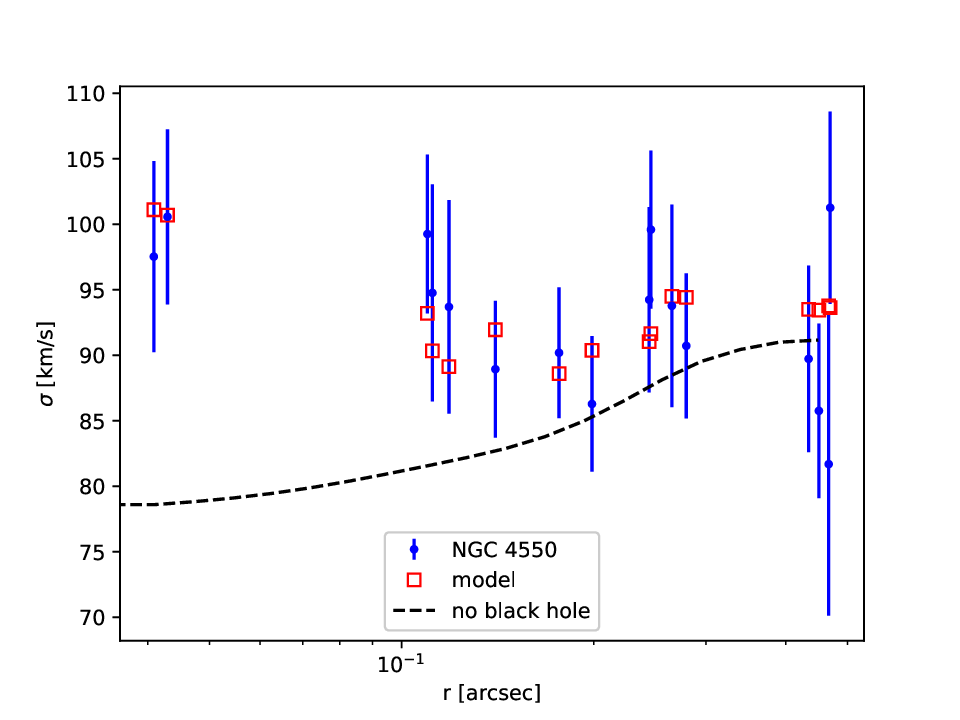}
\includegraphics[scale = 0.5]{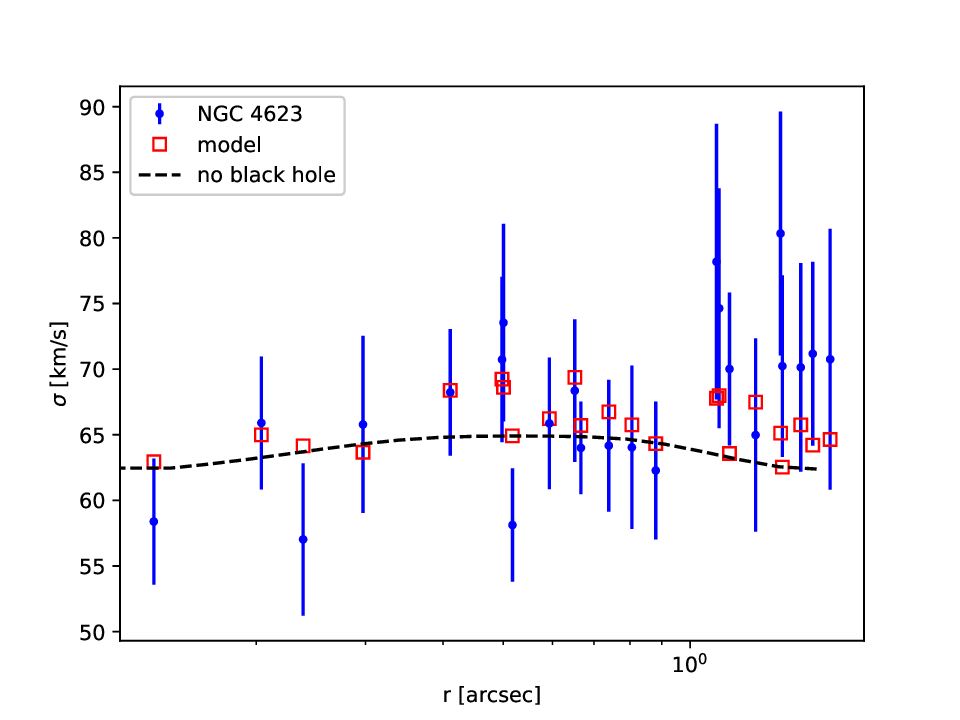}



 \caption{\label{fig:radkin} {\chng Comparison of the observed kinematics with the best-fitting JAM models. The best fitting JAM model with a black hole is calculated for the same Voronoi bins as the observations and both are then plotted as a function of projected radius from the centre of the galaxy. Also shown is a JAM model for the same galaxy without a black hole. In all galaxies except NGC 4623 the model without a black hole under-predicts the central velocity dispersion.}}
\end{center}
\end{figure*}

We show the radial distributions of velocity dispersion for all the galaxies in Fig. \ref{fig:radkin}. The figure also shows the {\chng predictions of the best fitting JAM model, as well as the no black hole case (as described in Sec.~\ref{sec:dynmodel})}. Galaxies NGC 4458, 4379, 4550 and 4262 show an increased velocity dispersion near the centre of the galaxy, indicative of the presence of an SMBH. The JAM models only show this rising velocity dispersion if a SMBH is included, whereas the models without a SMBH have decreasing velocity dispersions towards the centre. By contrast, the velocity dispersion of galaxy NGC 4623 decreases towards the centre of the galaxy. This is only consistent with the JAM model without a black hole.

{\chng As noted in Sec.~\ref{sec:dynmodel} we restricted the MGE model of NGC 4262 to near-circular components. When this constraint was removed, the galaxy inclination was forced to be $i>57$ degrees. This gave a higher black hole mass but a worse fit, over-predicting the central velocity dispersion by $3\sigma$.}

{\chng We present the posterior distributions of the four fitted parameters for the JAM models in Appendix \ref{sec:JAMresults}. As can be seen, there is some degeneracy in the models between the anisotropy of the galaxy, the mass-to-light ratio and the black hole mass.} {\chng Inclination is not well constrained (with the exception of NGC 4262), but has no major effect on the resulting black hole masses.} {\chng Despite the degeneracy between parameters, the models for the galaxies NGC 4262, 4379, 4458, and 4550 exclude the no black hole case at $3\sigma$, thus these galaxies have SMBH detections with 99.7\% certainty. The map for NGC4623 instead provides an upper limit with a no SMBH detection. }



We show the best fitting parameters of our JAM models in Table~\ref{tab:ownbh}. We are confident that these are good fits as the minimum reduced $\chi^2$ values were not large (of order unity) and the predicted velocity dispersion curves followed the shape of the observed curves well as shown in Fig. \ref{fig:radkin}. We calculated galaxy masses from the JAM model mass-to-light ratios; we also calculated the corresponding bulge masses using bulge-to-total ratios from \citet{Kraj2013a}. {\chng The inclination of NGC 4262 is more tightly constrained than for the other galaxies (see Table~\ref{tab:ownbh} and Fig~\ref{fig_JAM4262}). This is because the corresponding physical JAM parameter \citep[$q$, the deprojected axial ratio, see equation 14 of][]{JAMs} varies very rapidly with inclination when inclination is low, such as for face-on galaxies like NGC 4262. The 1 per cent uncertainty in inclination for NGC 4262 corresponds to 30 per cent uncertainty in $q$, whereas the 15 percent inclination uncertainty for NGC 4379 corresponds to an uncertainty of 18 per cent in $q$.}

With our observations from this paper, we have black hole measurements (or upper limits) for 11 of the 18 galaxies in our sample listed in Table~\ref{tab:sample}. We estimated a conservative lower limit for the average SMBH mass in our sample by assuming the unobserved galaxies do not contain any SMBHs. Our lower limit, the sum of the 11 detected black hole masses divided by the sample size of 18, is {\chng $3.7 \times 10^{7} M_\odot$.} This lower limit for the average SMBH mass is for an average total galaxy mass of $(1.81 \pm 0.14)\times 10^{10} M_\odot $ and an average bulge mass of $(1.31 \pm 0.15) \times 10^{10} M_\odot$.





\begin{table*}
\caption{\label{tab:ownbh} Supermassive black hole mass estimates}


 {\chng
 \begin{tabular}{l|r|r|r|r|r|r|r|r|r}
      {Galaxy} & Distance &Inclination &$\beta_z$ & {Black Hole Mass} & {M/L} &$\chi^2/DoF$ & $DoF$  &  Galaxy Mass & Bulge Mass\\
      NGC & Mpc & deg& & $M_\odot$ & $M_\odot / L_\odot$ & & &$M_\odot$ &$M_\odot$ \\
     \hline
      
4262&15.4 &  $20.3\pm  0.2 $ &  $0.511\pm 0.078 $ & $(3.15\pm 0.41) \times 10^{8}$ & $2.84\pm 0.18 $  & 0.48 & 109& $2.70 \times 10^{10}$ & $2.70 \times 10^{10}$  \\ 
4379&15.8 &  $81.5\pm 12.1 $ & $-0.058\pm 0.054 $ & $(7.07\pm 0.36) \times 10^{7}$ & $2.54\pm 0.07 $  &  2.4 & 90 & $2.21 \times 10^{10}$ & $2.21 \times 10^{10}$  \\ 
4458&16.4 &  $89.7\pm  7.9 $ &  $0.220\pm 0.040 $ & $(6.51\pm 0.45) \times 10^{7}$ & $1.60\pm 0.05 $  &  1.7 & 67 & $1.11 \times 10^{10}$ & $7.97 \times 10^{9}$  \\ 
4550&15.5 &  $74.5\pm  5.8 $ &  $0.274\pm 0.125 $ & $(1.14\pm 0.37) \times 10^{7}$ & $2.34\pm 0.23 $  & 0.76 & 12 & $2.08 \times 10^{10}$ & $1.31 \times 10^{10}$  \\ 
4623&17.4 &  $89.9\pm  4.6 $ &  $0.317\pm 0.092 $ & $ < 4.6 \times 10^{6}$ & $1.90\pm 0.16 $  & 0.78 & 19 & $1.04 \times 10^{10}$ & $2.19 \times 10^{9}$  \\ 
\\
\end{tabular}

 Notes: Distances are from \citet[][]{atlas3dfirst}. The uncertainties are $\pm 1 \sigma$ ranges. The galaxy masses were calculated from the MGE luminosities and the fitted mass to light ratios. The bulge masses were calculated from the galaxy masses using bulge-to-total ratios from \citet{Kraj2013a}. }

\end{table*}





\section{Discussion}

\begin{figure*}
\begin{center}
\includegraphics[scale = 0.5] {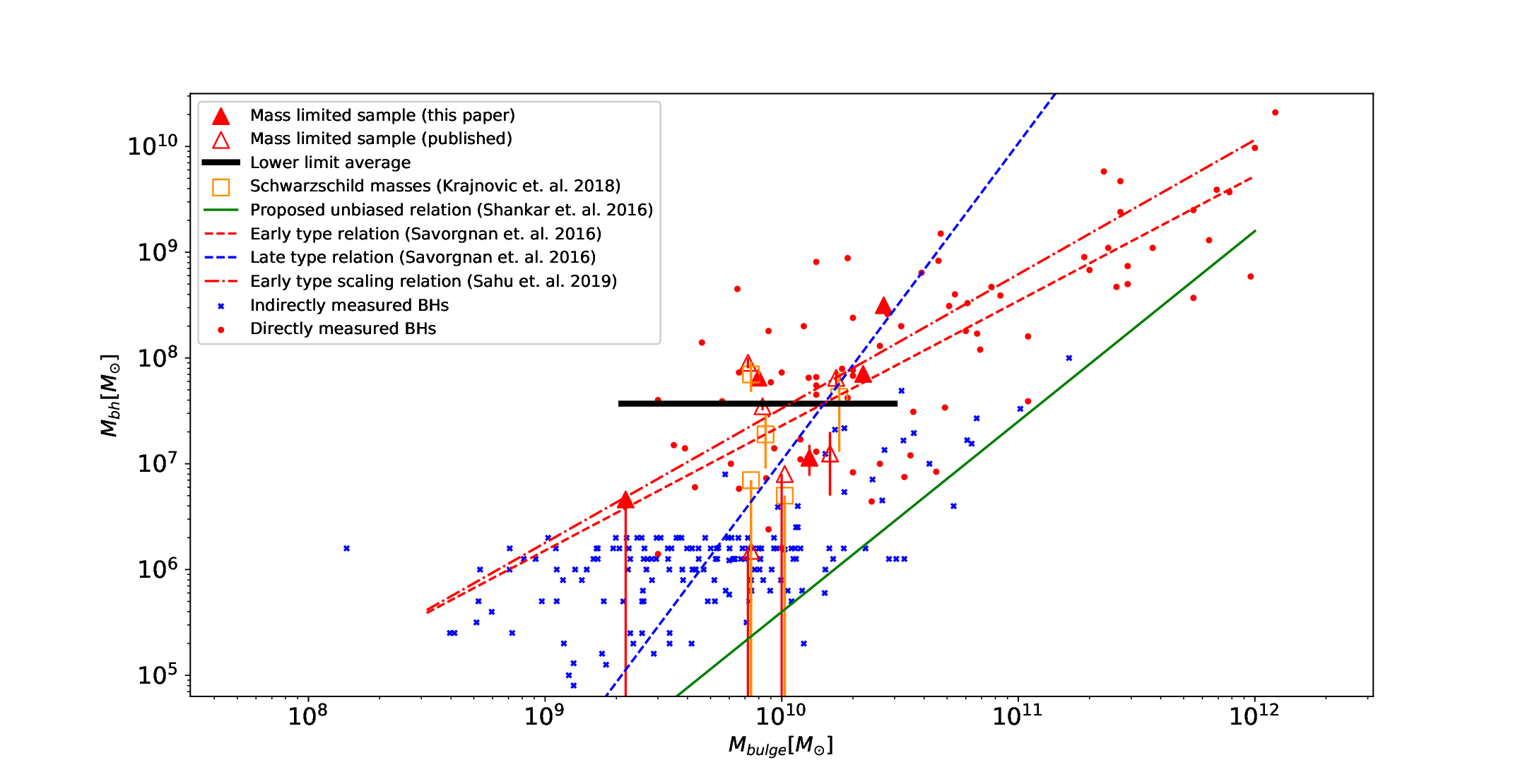}


 \caption{\label{fig_shankar} {\chng Comparison of black hole measurements to published scaling relations. Our mass-limited sample is shown as red triangles, including upper limits (uncertainties extend to the horizontal axis). The black line  shows the lower limit of the average black hole mass in our sample; this is not consistent with the `unbiased' relation (solid green line) proposed by \citet[][]{shankar16}. The small symbols are directly measured black holes \citep[red:][]{scott13,Graham_2014} and indirect black hole measurements \citep[blue:][]{Jiang11}. We also show alternative measurements for five galaxies using the Schwarzschild method \citep[orange squares:][]{krajonvicbh}}}
 
 \end{center}
\end{figure*}

In this section we first {\chng discuss the differences in axisymmetric Jeans models and triaxial Schwarzschild orbital models. We then }discuss the reliability of our measurements, before comparing our work to previous results, as summarised in Fig.~\ref{fig_shankar}. In particular, our mass-selected sample allows us to test for biases in the resulting scaling relations.

{\chng 
\subsection{Dynamical analysis method}
We used a single analysis method, Jeans Anisotropic Models (JAM), to estimate our black hole masses. This produced results similar to published black hole measurements using similar methods (see Fig~\ref{fig_shankar}). The JAM method is a common choice in the field, but more general approaches such Schwarzschild orbital modelling are also used. Schwarzschild modelling, as formulated by \citet[][]{Schwarzschild1979}, uses an orbital library, adjusting individual weights in a numerical superposition of orbits, to model the mass distribution of a galaxy \citep[][]{Thomas2010}. Further work extended the modelling to rotating triaxial galaxies \citep[][]{Schwarzschild1982}, and the methodology to compare to stellar kinematics \citep[][]{Bosch2008}. 

The Jeans models we use in this paper  restrict the allowed orbits, in particular, to match an axisymmetric mass potential of the galaxy. Due to these additional assumptions, the JAM methods give smaller uncertainties than the more general Schwarzschild approach \citep[e.g. fig.\ 10 of][]{denBrok2021}.

The increased generality of Schwarzschild orbital modelling over JAM models allows for broader application, although at the cost of computational time and the requirement of wider field kinematic data \citep[][]{Jin2019}. In particular, Schwarzschild models are more suitable for galaxies that do not satisfy the Jeans method assumptions, such as slow rotating, or misaligned galaxies where the rotational axes do not match the galaxy's structural axes. 

Some studies have found systematic offsets between black hole masses when comparing results from Jeans models and Schwarzschild orbital models. \citet[][]{krajonvicbh} and \citet[][]{Pilawa2022} found that JAM models appear to overestimate black hole masses when compared to triaxial Schwarzschild models, in the worst case, by a factor of two (average factor of 1.7; see Fig.~\ref{fig_shankar}), whereas \citet[][]{leung2018} found a difference of roughly 20\% over Schwarzschild models. In regards to the results of this paper, even the largest measured offset does not change the scope of our results, and would likely be covered by our measured uncertainties for most of our galaxies. However, there are galaxies within our sample that are not fully axisymmetric as assumed by the Jeans models: NGC 4458 and 4550 are both slow rotators, and additionally, NGC 4458 is misaligned. As such, we plan to assess our galaxies using the Schwarzschild method once we have observed our full sample.}



\subsection{Reliability of black hole mass measurements}
Here we describe some additional tests of our methodology.


We compared our model mass-to-light ratios to the expected values for stellar populations.  We used the \textsc{fsps} stellar population model \citep[][]{Conroy2009,Conroy2010}, the Miles spectral library \citep[][]{Sanchez2006}, Padova isochrones  \citep[][]{Girardi_et_al_00,MarigoGirardi2007,Marigo_et_al_08}, assuming a single stellar population and \citet[][]{Chabrier_03} initial stellar mass function. This model predicts mass-to-light ratios of 1--2 $M_\odot/L_\odot$ in our F850LP filter, unless the stellar population were old and very metal rich or alpha-enhanced. Our fitted values (1.4--2.5 $M_\odot/L_\odot$; see Table~\ref{tab:ownbh}) are consistent with these predictions. As a further test, our total galaxy masses--which depend on the mass-to-light ratios--are consistent with the independently-measured ATLAS 3D values. We do note that we assumed a constant mass-to-light ratio in our models: we believe this is reasonable as our kinematic data is tied to a very small region in each galaxy. 


The influence of dark matter on the supermassive black holes is contested. As discussed in studies like that by \citet[][]{schulzedarkmatter}, the presence of dark matter in the centre of a galaxy can cause models to misrepresent the mass-to-light ratio, and thus overestimate the stellar mass component of the galaxy. Both \citet[][]{ruslidarkmatter} and \citet[][]{schulzedarkmatter} found that the presence of concentrated dark matter in high velocity dispersion, early-type galaxies can cause dynamical models to underestimate black holes by a factor of 1.5 to 2. We present our results under the assumption that our galaxies do not have a high concentrations of dark matter in their central regions. This assumption is supported by simulations of early type galaxies where baryonic matter dominates dark matter within one effective radius of the centre of the galaxy \citep[][]{Xu2017}. Additionally, as discussed by \citet[][]{atlas3d}, the JAM methodology uses observed kinematics to estimate the enclosed masses. This enclosed mass should include any dark matter present.

\subsection{Black hole scaling relations}


Our mass-limited sample of 18 galaxies (with 11 black hole measurements to date) is too small to define a new scaling relation, but we have calculated a lower limit to the average black hole mass for our sample of {\chng $3.7 \times 10^{7} M_\odot$ }at an average galaxy bulge mass of $(1.31 \pm 0.15) \times 10^{10} M_\odot$. We can compare this to other scaling relations at the same galaxy mass, as well as testing the hypothesis that some of the published relations are biased. We give details of the various published scaling relations we refer to in Appendix~\ref{sec:relation-formulae}.

\subsubsection{The effect of possible bias}

As we describe in Sec.~\ref{sec:intro}, \citet[][]{shankar16} suggest that some published scaling relations are strongly biased by the difficulty of detecting SMBHs in smaller galaxies so that only the most massive SMBHs are detected in these galaxies. It is certainly true that some published scaling relations do exclude non-detections of black holes from the calculation \citep[e.g. ][]{krajonvicbh} but the amount of bias has not been tested empirically. \citet[][]{shankar16} suggest the true scaling relations are much steeper than the published relations, proposing three models of unbiased scaling relations. We show the first of their proposed unbiased relations in Fig.~\ref{fig_shankar} (see Appendix~\ref{sec:relation-formulae} for details).

The aim of our mass-limited sample is to test for this bias as we will measure all galaxies in a sample for black holes, including non-detections. We have not yet observed the full sample, but our lower limit, as shown by the horizontal line in Fig.~\ref{fig_shankar}, is already clearly higher than the proposed unbiased relation from \citet[][]{shankar16} which predicts black hole masses of $6 \times 10^{5} M_\odot$ at the galaxy mass of $1.31 \times 10^{10} M_\odot$ where our lower limit for the complete sample is a black hole mass of {\chng $3.7 \times 10^{7} M_\odot$.} Our value is $5\sigma$ greater than the proposed unbiased prediction in terms of the expected 0.25 dex scatter in that relation \citep[Appendix~\ref{sec:relation-formulae}, ][]{shankar16}, suggesting that the bias is much less than suggested. {\chng If we had used the Schwarzchild modelling approach we might expect our masses to be 40 per cent (0.23 dex) lower \citep[][]{krajonvicbh} but we would still exclude the relation at  $4\sigma$.} When our sample is complete we will be able to estimate the actual bias. 

In seeking to explain the very strong bias predicted by \citet[][]{shankar16}, two other effects may be contributing. As discussed by \citet[][]{Kormendy_2013}, \citet[][]{heckman2014} and \citet[][]{Shankar_2019} and in Sec.~\ref{sec:intro}, measured black hole mass heavily depends on the type(s) of galaxy measured and the methodology of measurement. Our sample is specifically limited to early type galaxies, whereas the \citet[][]{shankar16} analysis includes a range of galaxy types, in particular the late type galaxies (the active galactic nucleus or AGN sample in that paper). AGN samples follow a much steeper scaling relation and have black hole masses more than an order of magnitude lower than early type galaxies at the same galaxy mass \citep[e.g.\ Fig.~\ref{fig_shankar} and ][]{Savorgnan2016,Shankar_2019}. Black hole masses in the AGN galaxy samples are calculated with a different methodology to the direct dynamical modelling we used for early type galaxies: they use a virial estimator based on the broad emission line widths \citep[e.g.\ ][]{Reines2013}. Our results suggest that the large difference between the black hole masses in the AGN and early-type galaxies is not due to a large bias in the methodlogy of published measurements for early type galaxies, but is  more likely due to the different physical processes for the different types of galaxies.



\subsubsection{Comparison with published scaling relations}

We now compare our results with some published black hole scaling relations. As noted above, our sample is not yet fully observed so our analysis focuses on consistency checks. We also note that our approach, to find the average black hole mass in a complete galaxy sample, is different to that often used in creating scaling relations where galaxies without black hole detections are not included in the calculations.

We first consider two recent scaling relations for early type galaxies \citep[][]{Savorgnan2016,Sahu19}. These are shown in Fig.~\ref{fig_shankar} with details in Appendix~\ref{sec:relation-formulae}. The first relation \citep[][]{Savorgnan2016} is consistent with our sample, passing through our detected black hole masses and slightly above our lower limit to the average black hole mass at our mean galaxy mass of $(1.31 \pm 0.15) \times 10^{10} M_\odot$. The second relation \citet[][]{Sahu19} is slightly below our lower limit for the average black hole mass, but not significantly so considering its scatter of 0.6 dex. Even when our sample is completely measured, we might still expect our average mass to lie below these published relations given that we include the galaxies without detected black holes.

\begin{figure}
\begin{center}
\includegraphics[scale = 0.5] {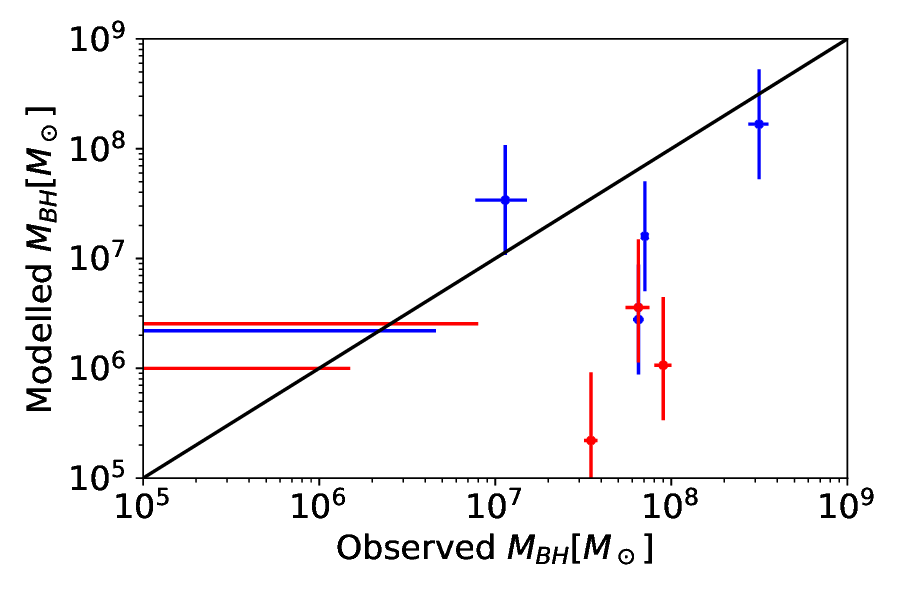}

 \caption{ \label{fig_kra} The black hole multi-variable model proposed by \citet[][]{Kraj2017} compared to the observed black hole masses in our sample, from this paper (blue) and \citet[][]{krajonvicbh} (red). Upper limits are shown as horizontal lines. The vertical error bars give the expected 0.5 dex scatter in the model masses. The black diagonal line denotes equality between the observed and modelled black hole masses. }

 \end{center}
\end{figure}

We also compare our results to a multi-variable relation predicting black hole mass from both the mass and radius of the host galaxy \citep[][]{Kraj2017}. As this depends on two variables we present this, in Fig.~\ref{fig_kra}, as a comparison between the directly measured black hole masses from our sample and the black hole masses predicted by the multi-variable model.  We describe the relation and our calculation of model black hole masses in Appendix~\ref{sec:relation-formulae}. Fig.~\ref{fig_kra} shows that while all of our direct measurements are within $2\sigma$ of the expected  scatter of the model black hole masses, there is a trend of our detected masses to be higher than the model on average. We cannot confirm this until our measurements are more complete, but it may reflect the fact our sample is limited to early type galaxies whereas the \citet[][]{Kraj2017} model was based on a wide range of galaxy types.


\section{Conclusion}

In this paper we present black hole measurements for 5 early type galaxies (NGC 4262, NGC 4379, NGC 4458, NGC 4550 and NGC 4623) using the NIFS and Osiris instruments from the Keck and Gemini telescopes respectively. Our aim is to obtain black hole masses in a mass-selected galaxy sample to test claims of bias in black hole scaling relations. Our results are follows.
\begin{itemize}
    \item We detect and measure black hole masses in 4 of the 5 galaxies we observed (Table~\ref{tab:sample}).
    \item 11 of the 18 galaxies in our mass-limited sample have now been measured for black holes.
    \item We calculate a lower limit for the average black hole mass in our sample of {\chng $3.7 \times 10^{7} M_\odot$.} This is at an average galaxy mass of $(1.81 \pm 0.14)\times 10^{10} M_\odot $ and an average bulge mass of $(1.31 \pm 0.15) \times 10^{10} M_\odot$.
    \item Our lower limit shows that black hole masses in early type galaxies are not strongly affected by the detection biases discussed by \citet[][]{shankar16}.
    \item Our initial measurements are consistent with current black hole scaling relations for early type galaxies, although we expect our final results to be lower as we include non-detections. 
    \item {\chng We plan to re-evaluate the measured black hole masses using Schwarzschild modelling once we have observed the full sample in order to avoid the assumptions in axisymmetric Jeans models.}
\end{itemize}

\section*{Acknowledgements}

We are very grateful to the staff of the Gemini-North and the Keck telescopes for their assistance and support with the observation runs GN-2016A-Q-35, GN-2017A-Q-22 and Z039OL. We appreciate the assistance of Eugene Vasiliev in providing advice about the Schwarzschild orbital method. We thank the referee of this paper for advice which has greatly improved this work.

JP is supported by the Australian government through the Australian Research Council's Discovery Projects funding scheme (DP200102574). 

CS acknowledges support from ANID/CONICYT through FONDECYT Postdoctoral Fellowship Project No. 3200959. 

SMS acknowledges funding from the Australian Research Council (DE220100003).

SM is supported by the Senior Research Scientist (SRS) grant no. SB/SRS/2020-21/56/PS by the Science Engineering and Research Board (SERB) of the Govt. of India.

Parts of this research were conducted by the Australian Research Council Centre of Excellence for All Sky Astrophysics in 3 Dimensions (ASTRO 3D), through project number CE170100013.
The data we used from the ACSVCS survey was supported through a grant from the Space Telescope Science Institute, which is operated by the Association of Universities for Research in Astronomy, Inc., under NASA contract NAS5-2655.


\section*{Data Availability}

The data for this paper were provided by the Gemini and W. M. Keck observatories. Details on the observing programs can be found at \url{https://archive.gemini.edu/searchform/} for our Gemini observing runs GN-2016A-Q-35 and GN-2017A-Q-22, and \url{https://www2.keck.hawaii.edu/koa/public/koa.php} for our Keck program Z039OL. The derived data generated in this research will be shared on reasonable request to the corresponding author.



\bibliographystyle{mnras}
\bibliography{example} 




\appendix

\section{Jeans Anisotropic Model Results}
\label{sec:JAMresults}


\begin{figure*}
\begin{center}
\includegraphics[scale = 0.7]{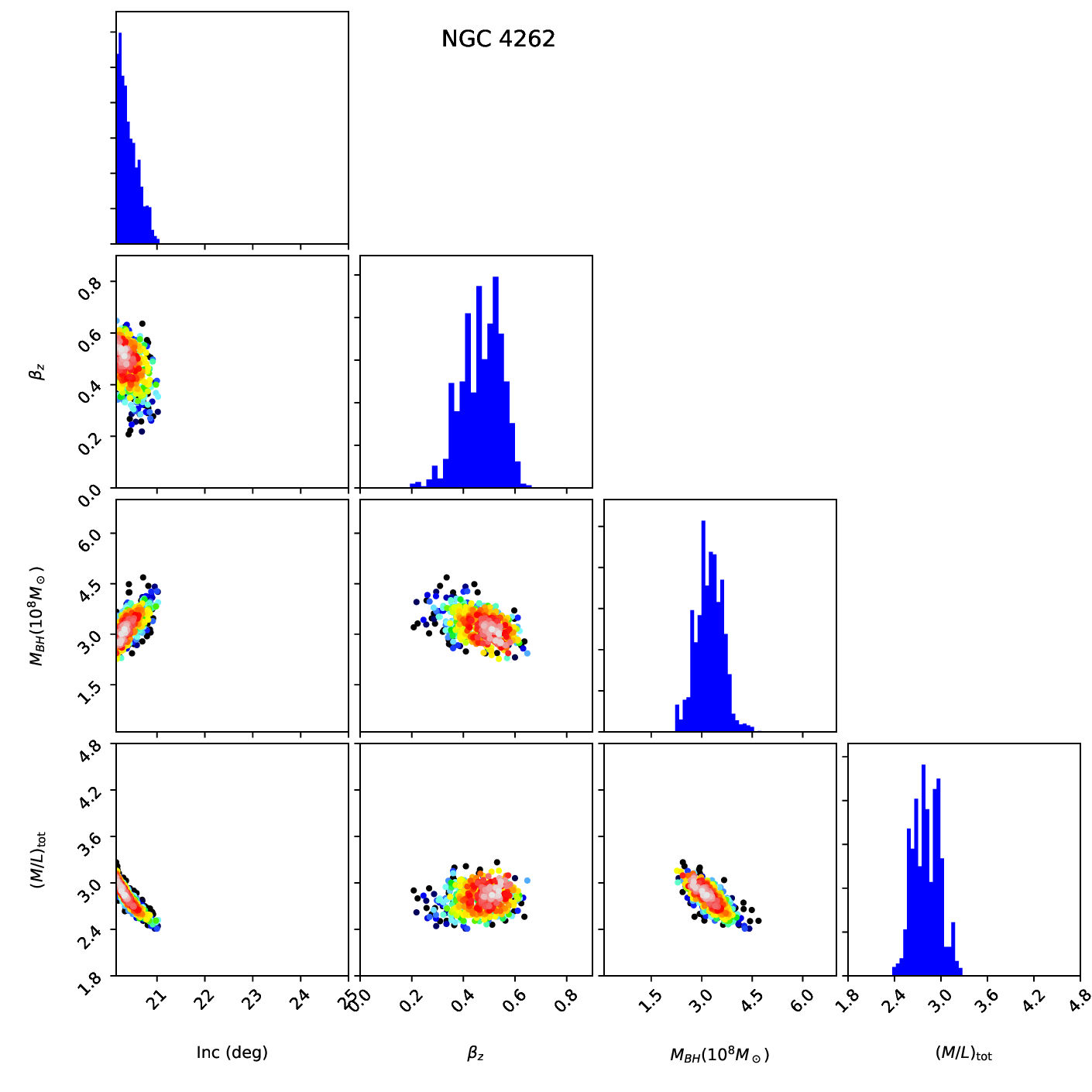}
 \caption{ \label{fig_JAM4262} Posterior distributions of the JAM parameters for galaxy NGC 4262. The two-parameter plots are coloured according to log-likelihood with white as the most likely and black points being 3 sigma or more away from the best fit. }

 \end{center}
\end{figure*}

\begin{figure*}
\begin{center}
\includegraphics[scale = 0.7]{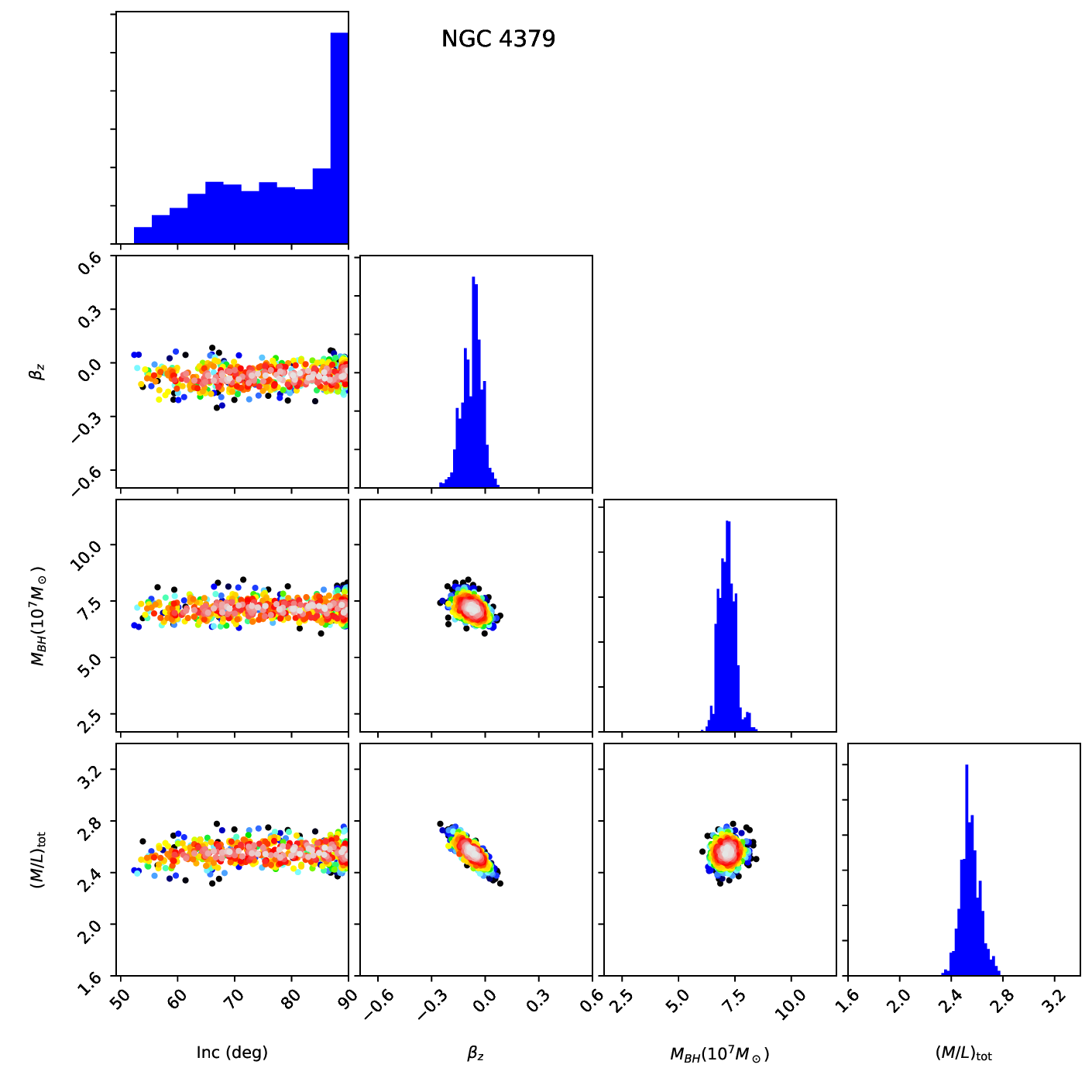}
 \caption{ \label{fig_JAM44379} Posterior distributions of the JAM parameters for galaxy NGC 4379. The two-parameter plots are coloured according to log-likelihood with white as the most likely and black points being 3 sigma or more away from the best fit. }

 \end{center}
\end{figure*}

\begin{figure*}
\begin{center}
\includegraphics[scale = 0.7]{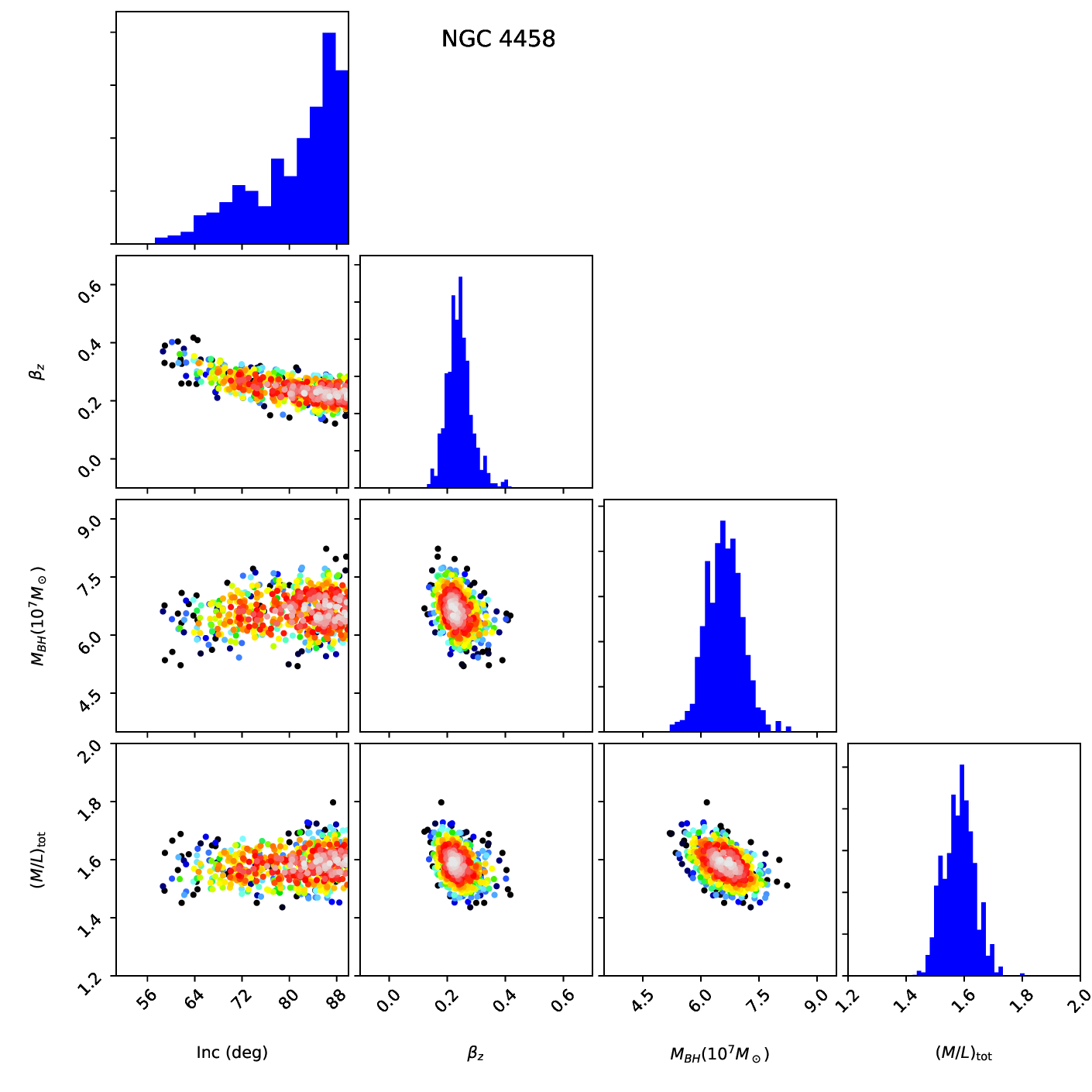}
 \caption{ \label{fig_JAM4458} Posterior distributions of the JAM parameters for galaxy NGC 4458. The two-parameter plots are coloured according to log-likelihood with white as the most likely and black points being 3 sigma or more away from the best fit. }

 \end{center}
\end{figure*}

\begin{figure*}
\begin{center}
\includegraphics[scale = 0.7]{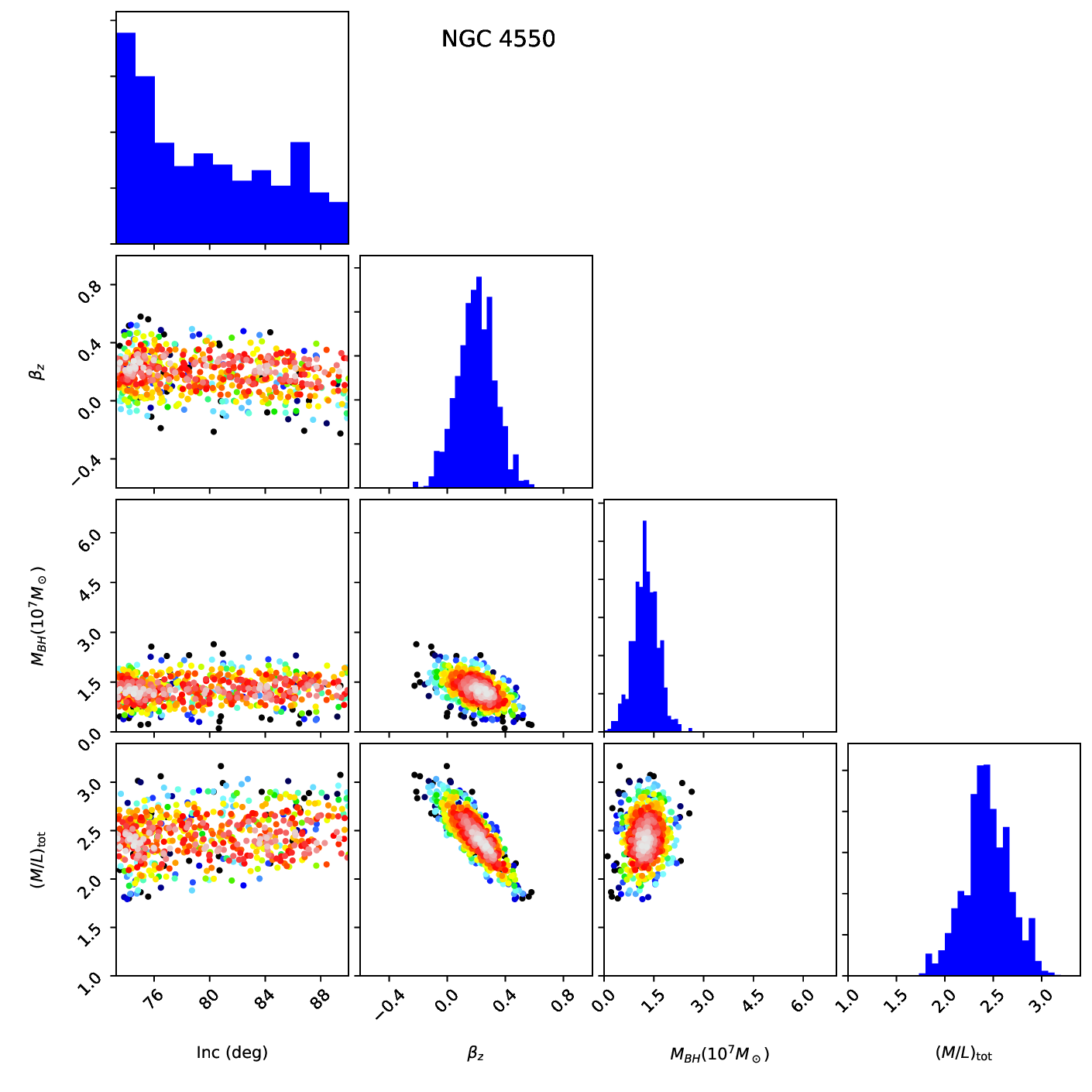}
 \caption{ \label{fig_JAM4550} Posterior distributions of the JAM parameters for galaxy NGC 4550. The two-parameter plots are coloured according to log-likelihood with white as the most likely and black points being 3 sigma or more away from the best fit. }

 \end{center}
\end{figure*}

\begin{figure*}
\begin{center}
\includegraphics[scale = 0.7]{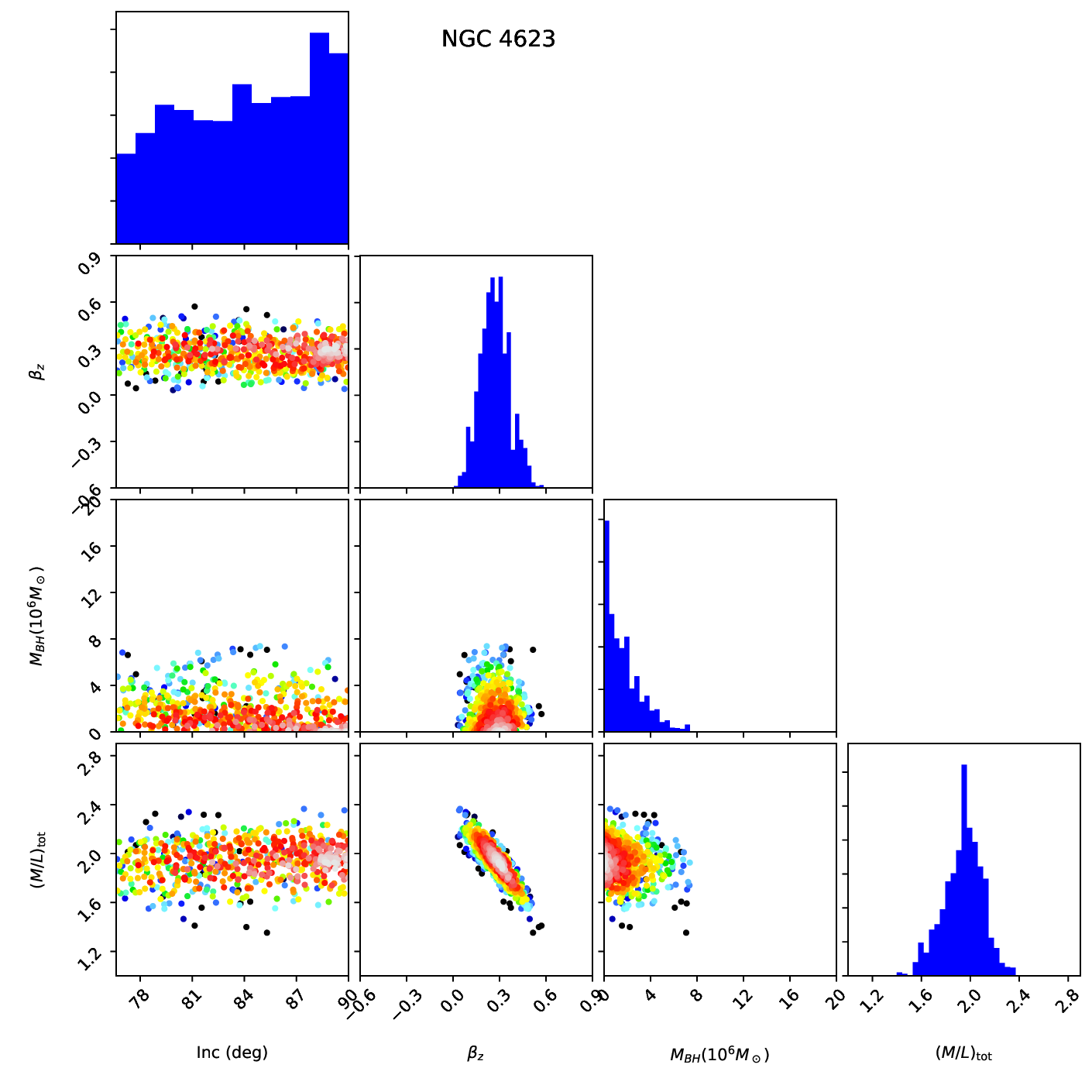}
 \caption{ \label{fig_JAM4623} Posterior distributions of the JAM parameters for galaxy NGC 4623. The two-parameter plots are coloured according to log-likelihood with white as the most likely and black points being 3 sigma or more away from the best fit.  }

 \end{center}
\end{figure*}

\section{Published scaling relations}
\label{sec:relation-formulae}

\subsection{Proposed unbiased relation \citep[][]{shankar16}}

\citet[][]{shankar16} proposed three different models of unbiased scaling relations that would match published scaling relations based on dynamical measurements after a selection bias was applied. We have chosen their ``Model I'' as it gave the best match to direct black hole observations when a biased selection was applied. It has the following form:
\begin{equation}
    \log\frac{M_{BH}}{M_\odot} = \gamma + \beta \log(\frac{\sigma}{200 kms^{-1}}) + \alpha \log(\frac{M_{*}}{10^{11} M_\odot}),    
\end{equation}
with $\gamma=7.7$, $\beta=4.5$, and $\alpha=0.5$ \citep[][]{shankar16}. We assumed a constant value of $\sigma = 160 kms^{-1}$ for the galaxies in the lmimited mass range of our sample. The scatter in predicted $\log(M_{BH})$ is 0.25 dex.

\subsection{Early and late type relations \citep[][]{Savorgnan2016}}


\citet[][]{Savorgnan2016} fitted scaling relations to samples of 45 early-type galaxies and 17 late type galaxies with the form
\begin{equation}
    \log(\frac{M_{BH}}{M_\odot}) = \alpha + \beta \log(\frac{M_{bulge}}{\langle M_{bulge}\rangle }).
\end{equation}
For the early type sample $\alpha=8.56\pm0.07$, $\beta=1.04\pm0.10$, and $\langle M_{bulge}\rangle = 10^{10.81}$ with an rms scatter in $\log(M_{BH})$ of 0.5 dex.
For the late type sample $\alpha=7.18\pm0.21$, $\beta=3.0\pm1.3$, and $\langle M_{bulge}\rangle = 10^{10.05}$ with an rms scatter $\log(M_{BH})$ of 0.9 dex.

\subsection{Early type relation \citep[][]{Sahu19}}

\citet[][]{Sahu19} fitted a scaling relation to a sample of 84 early-type galaxies (ETGs) with directly measured supermassive black hole masses with this form:
\begin{equation}
\log(M_{BH}) = (8.41 \pm{0.06})  + (1.27 \pm{0.07}) \log(\frac{M_{*,sph}}{v(5\times10^{10} M_\odot)}),
\end{equation}
where $v$ is a conversion coefficient for galaxy masses calculated with a filter-specific mass-to-light ratio \citep[][]{Sahu19}. We used $v=1$ as our masses used mass-to-light ratios calculated directly for each galaxy (see Sec~\ref{sec:samplesel}).The scatter in predicted $\log(M_{BH})$ is 0.6 dex.

\subsection{Multi-variate relation \citep[][]{Kraj2017}}

\citet[][]{Kraj2017} analysed a large sample of 181 galaxies with detected black holes to fit two-variable scaling relations of this form:
\begin{equation}
    \log(\frac{M_{BH}}{10^{8} M_\odot}) = a + b \log(\frac{M_{*}}{10^{11} M_\odot}) + c\log(\frac{R_e}{5kpc}).
\end{equation}
\citet[][]{Kraj2017} used two different approaches to solve this equation. There were minimal differences in the black hole estimates between the two; we have adopted the first with values of $a=7.66\pm0.06$, $b=2.7\pm0.2$, and $c=-2.9\pm0.3$. The scatter in predicted $\log(M_{BH})$ is 0.5 dex.

We used the following procedure to apply this scaling relation to our sample (both published black hole measurements and the results of this paper) consistently with the method used by \citet[][]{Kraj2017}. First, we estimated galaxy stellar masses from Two Micron All-Sky Survey photometry, as per equation 2 of \citet[][]{Cappellari2013}:
\begin{equation}
    \log(M_{*}) = 10.58 - 0.44 \times (M_{K_s}+23).
\end{equation}
We then used galaxy effective radii directly from  the ATLAS 3D survey \citep[][]{atlas3d}.


\bsp	
\label{lastpage}
\end{document}